\documentclass[11pt]{article}

\usepackage{subcaption}
\usepackage{xcolor}
\usepackage{microtype}
\usepackage{url}
\usepackage{amsfonts} 
\usepackage{nicefrac}
\usepackage{mathtools, nccmath}
\usepackage{minitoc}
\usepackage{cancel}
\usepackage{microtype}
\usepackage{graphicx}
\usepackage{booktabs} 
\usepackage{comment}
\usepackage{url}
\usepackage{amsmath}
\usepackage{amsthm}
\usepackage{amscd}
\usepackage{t1enc}
\usepackage{enumerate}
\usepackage{indentfirst}
\usepackage{listings}
\usepackage{graphics}
\usepackage{pict2e}
\usepackage{epic}
\usepackage{epstopdf} 
\usepackage{minitoc}
\usepackage{bm}
\usepackage{wrapfig}
\usepackage{algorithm}
\usepackage{algorithmic}
\usepackage{environ}
\usepackage{tikz}
\usepackage{float}
\usepackage{paralist}
\usepackage{blindtext}
\usepackage{pgfplots}
\usepackage{tikz}
\usepackage{amssymb}
\usepackage{setspace}
\usepackage{xfrac}
\usepackage{xstring}
\usepackage{xspace}
\usepackage{microtype}
\usepackage{graphicx}
\usepackage{gensymb}
\usepackage{enumitem}
\usepackage{tabularx}
\usepackage{fancyhdr}
\usepackage{soul}
\usepackage[mathscr]{eucal}
\usepackage[hidelinks]{hyperref}
\usepackage[most]{tcolorbox}
\usepackage[left=1in, right =1in, top=1in, bottom=0.7in]{geometry}
\usepackage{lineno}

\newtheorem{definition}{Definition}
\theoremstyle{definition}

\pgfplotsset{compat=1.14}
\pdfoutput=1

\usepackage[style=numeric-comp,sorting=none]{biblatex}
\title{\textbf{When Are Combinations of Humans and AI Useful? - A Systematic Review and Meta-Analysis}}
\allowdisplaybreaks
\usepackage{authblk}

\renewcommand{\headrulewidth}{0pt}

\author[1,2,$^*$]{Michelle Vaccaro}
\author[1,$^*$]{Abdullah Almaatouq}
\author[1,$^*$]{Thomas Malone}
\affil[1]{MIT Center for Collective Intelligence, Sloan School of Management, Massachusetts Institute of Technology, Cambridge, USA}
\affil[2]{Institute for Data, Systems, and Society, Schwarzman College of Computing, Massachusetts Institute of Technology, Cambridge, USA}
\affil[$^*$]{Corresponding authors, vaccaro@mit.edu, amaatouq@mit.edu, and malone@mit.edu}

\date{}
\begin{document}
\newgeometry{left=1in, right =1in, top=0.6in, bottom=0.7in}

\begin{titlepage}
\maketitle
\vspace{-1cm}
\begin{abstract}

Inspired by the increasing use of AI to augment humans, researchers have studied human-AI systems involving different tasks, systems, and populations. Despite such a large body of work, we lack a broad conceptual understanding of when combinations of humans and AI are better than either alone.  Here, we addressed this question by conducting a pre-registered systematic review and meta-analysis of over 100 recent experimental studies reporting over 300 effect sizes. First, we found that, on average, human-AI combinations performed significantly \textit{worse} than the best of humans or AI alone (Hedges' $g = -0.23$, 95\% confidence interval $-0.39$ to $-0.07$). Second, we found performance \textit{losses} in tasks that involved making decisions and significantly greater \textit{gains} in tasks that involved creating content. Finally, when humans outperformed AI alone, we found performance \textit{gains} in the combination, but when the AI outperformed humans alone we found \textit{losses}. These findings highlight the heterogeneity of the effects of human-AI collaboration and point to promising avenues for improving human-AI systems.


\end{abstract}

\end{titlepage}
\restoregeometry

\section{Introduction}
People increasingly work with artificial intelligence (AI) tools in fields including medicine, finance, and law, as well as in daily activities such as traveling, shopping, and communicating.  These human-AI systems have tremendous potential given the complementary nature of humans and AI -- the general intelligence of humans allows us to reason about diverse problems, and the computational power of AI systems allows them to accomplish specific tasks that people find difficult.

In fact, a large body of work suggests that integrating human creativity, intuition, and contextual understanding with AI's speed, scalability, and analytical power can lead to innovative solutions and improved decision-making in areas such as healthcare~\cite{bohr2020rise},  customer service~\cite{nicolescu2022human}, and scientific research~\cite{koepnick2019novo}. On the other hand, a growing number of studies reveal that human-AI systems do not necessarily achieve better results when compared to the best of humans or AI alone.  Challenges such as communication barriers, trust issues, ethical concerns, and the need for effective coordination between humans and AI systems can hinder the collaborative process~\cite{bansal2021does, buccinca2020proxy, lai2020why, zhang2020effect, bansal2019updates, vaccaro2019the}. 

These seemingly contradictory results raise important questions: When do humans and AI complement each other? And by how much? To address these issues, we conducted a systematic literature review and meta-analysis in which we quantified synergy in human-AI systems and identified factors that explain its presence (or absence) in different settings.

We focused on two outcomes: (1) \textit{human-AI synergy}, where the human-AI group performs better than \textit{both} the human alone and the AI alone, which is analogous to strong synergy in human groups~\cite{larson2013search, almaatouq2021task}; and (2) \textit{human augmentation}, where the human-AI group performs better than the human alone (see Section S1.1 in the SI for more details).



When evaluating human-AI systems, many studies focus on human augmentation~\cite{bo2021toward, boskemper2022measuring, bondi2022role, schemmer2022meta}.  This measure can serve important purposes in contexts for which full automation cannot happen for legal, ethical, or safety reasons and in cases when AI does not align with human values. But when talking about the potential of human-AI systems, most people implicitly assume that the combined system should be better than either alone; otherwise, they would just use the best of the two~\cite{wilson2018collaborative}. Thus they are looking for human-AI synergy.  In light of these considerations, a growing body of work emphasizes evaluating and searching for synergy in human-AI systems~\cite{bansal2021does, bansal2021most, bansal2019updates, wilder2020learning, rastogi2023taxonomy, mozannar2024effective}.


To evaluate this synergy in human-AI systems, we analyzed 370 unique effect sizes from 106 different experiments published between January 2020 and June 2023 that included the performance of the human-only, AI-only, and human-AI systems. On average, we found evidence of human augmentation, meaning that the average human-AI systems performed better than the human alone. But we did not find human-AI synergy on average, meaning that the average human-AI systems performed \textit{worse} than at least one of the human alone or the AI alone. So, in practice, if we consider only the performance dimensions the researchers studied, it would have been better to use either a human alone or an AI system alone rather than the human-AI systems studied.    

While this overall result may appear discouraging, we also identified specific factors that did or did not contribute to synergy in human-AI systems.  On the one hand, for example, much of the recent research in human-AI collaboration has focused on using AI systems to help humans make decisions by providing not only suggested decisions but also confidence levels or explanations. But we found that neither of these factors significantly affected the performance of human-AI systems.

On the other hand, little work has investigated the effects of  task types and the relative performance of humans alone and AI alone. But we found that both factors significantly affected human-AI performance.  Our work thus sheds needed light on promising directions for designing future human-AI systems to unlock the potential for greater synergy.

\section{Results}\label{sec:results}

Our initial literature search yielded 5126 papers, and, per the review process described in the Methods (see section \ref{lit_review}), we identified 74 that met our inclusion criteria (see Figure S1).  These papers reported the results of 106 unique experiments, and many of the experiments had multiple conditions, so we collected a total of 370 unique effect sizes measuring the impact of human-AI collaboration on task performance. Figure S2 highlights the descriptive statistics for the effect sizes in our analysis.  We synthesize these data in a three-level meta-analytic model (see the Methods section \ref{data_analysis}).  We also make the materials required to reproduce our results publicly accessible through an Open Science Framework (OSF) repository.

\subsection{Overall Levels of Human-AI Synergy}\label{sec:main_results}

In our primary analyses, we focus on human-AI synergy, so we compare the performance of the human-AI systems to a baseline of the human alone or the AI alone, whichever performed best. We found that the human-AI systems performed significantly \textit{worse} overall than this baseline. The overall pooled effect was negative ($g = -0.23$, $t(92)=-2.89$, two-tailed $p = 0.005$, 95\% confidence interval (CI) $-0.39$ to $-0.07$) and considered small according to conventional interpretations~\cite{cohen2013statistical}.  

\begin{figure}[H]
  \begin{subfigure}[b]{0.45\textwidth}
     \includegraphics[width=\textwidth]{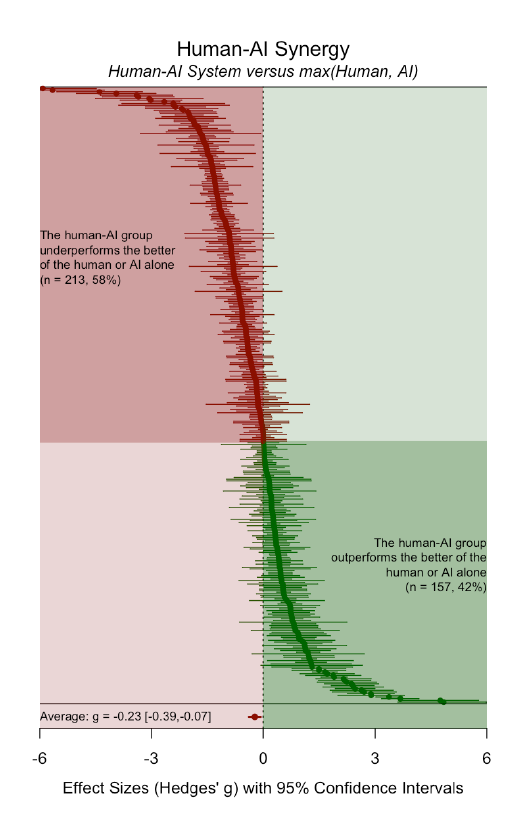}
    \caption{}
    \label{fig:forest_plot_strong}
  \end{subfigure}
    \hfill
  \begin{subfigure}[b]{0.45\textwidth}
    \includegraphics[width=\textwidth]{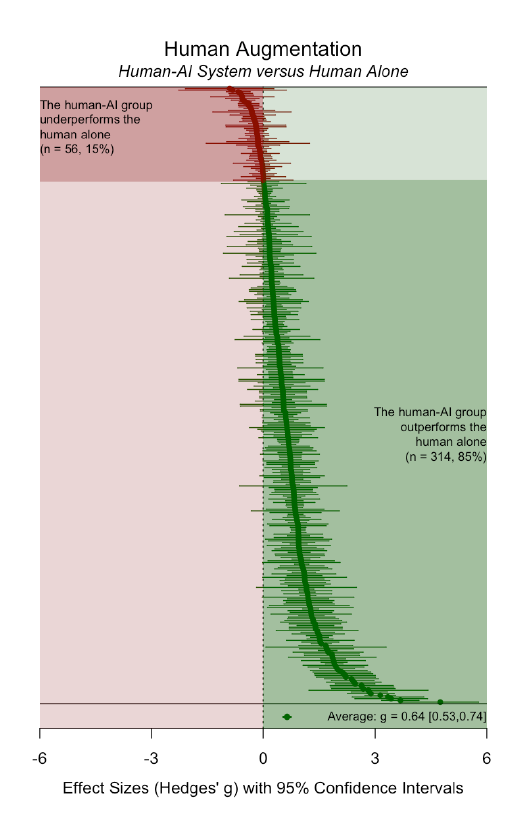}
    \caption{}
    \label{fig:forest_plot_human}
  \end{subfigure}
  \caption{
  Forest plots of all effect sizes ($k=370$) included in the meta-analysis.  The positions of the points on the $x$-axes represent the values of the effect sizes, and the bars indicate the 95\% confidence interval for the effect sizes.  The colors of the points and lines correspond to the values of the effect sizes, with negative effect sizes colored red and positive effect sizes colored green.  The black dotted line corresponds to an effect size of Hedges' $g = 0$, which means that the human-AI system performed the same as the baseline.  The circle at the bottom of the graph represents the meta-analytic average effect size and confidence interval.
  }
  \label{fig:forest_plot2}
\end{figure}

On the other hand, when we compared the performance of the human-AI systems to a different baseline -- the humans alone -- we found substantial evidence of human augmentation.  The human-AI systems performed significantly \textit{better} than humans alone, and this pooled effect size was positive ($g = 0.64$, $t(98)=11.87$, two-tailed $p = 0.000$, 95\% CI $0.53$ to $0.74$) and medium to large in size~\cite{cohen2013statistical}. Figure \ref{fig:forest_plot2} displays a forest plot of these effect sizes. In other words, the human-AI systems we analyzed were, on average, better than humans alone but not better than both humans alone and AI alone. For effect sizes that correspond to other potential outcomes of interest, see Table S3 and Figure S3.


\subsection{Heterogeneity of Human-AI Synergy}\label{sec:moderator_analysis}
\begin{figure}[H]
    \centering
    \includegraphics[width=.8\textwidth]{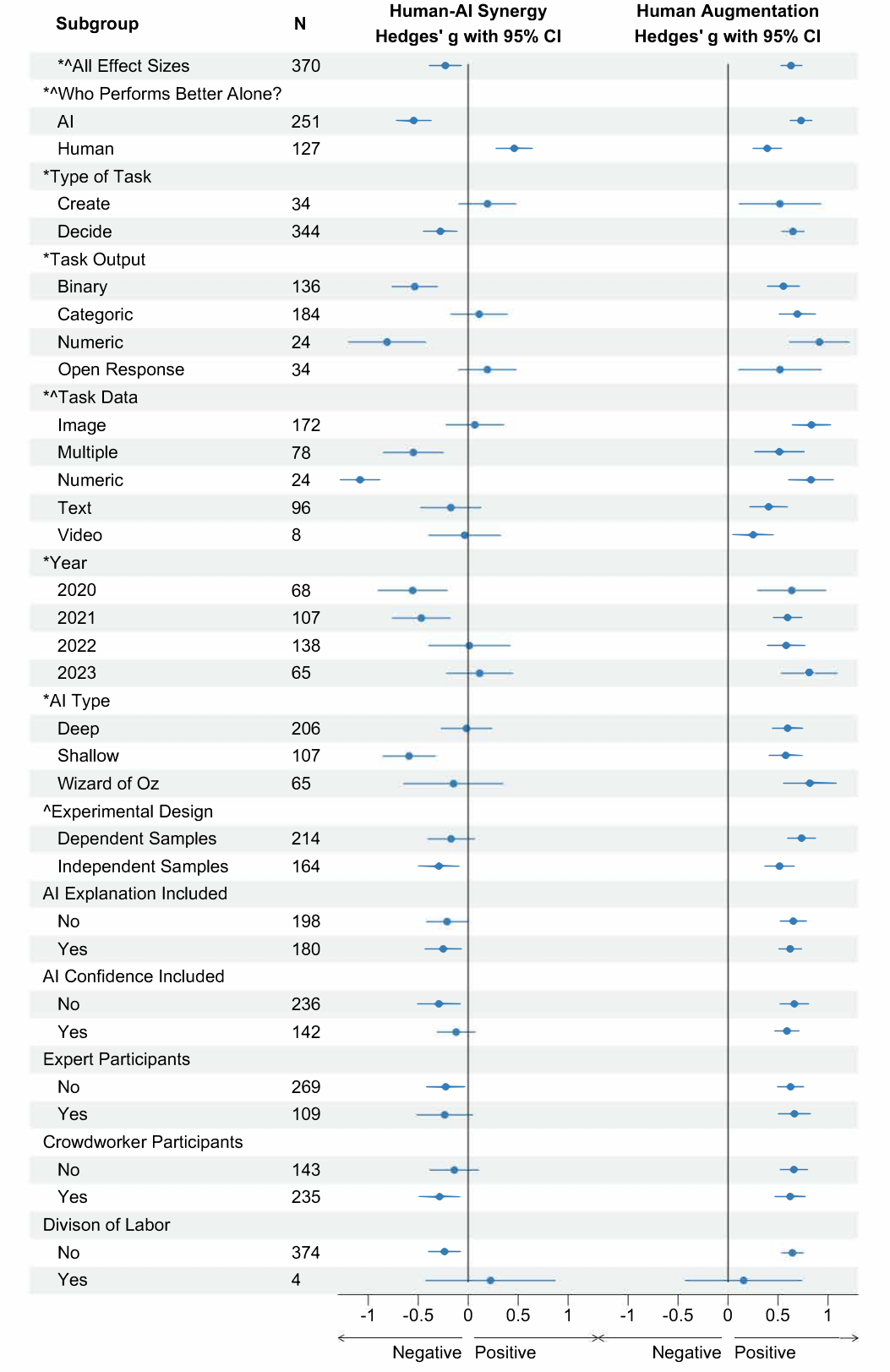}
  \caption{
  Results from the three-level meta-regression models for the moderator variables.  Here, $N$ is the number of included effect sizes for the moderator subgroup level, and the estimated effect size with the corresponding 95\% confidence interval. The symbols in front of the moderator indicate if there is a statistically significant difference between the subgroups for human-AI synergy (*) and human augmentation ($^{\wedge}$). 
  }
  \label{fig:mod_plot}
\end{figure}

We also found evidence for substantial heterogeneity of effect sizes in our estimations of human-AI synergy ($I^2=97.7\%$) and human augmentation ($I^2=93.8\%$) (see table Table S5 and S6 for more details). Our moderator analysis identified characteristics of participants, tasks, and experiments that led to different levels of human-AI synergy and human augmentation, and it helps explain sources of this heterogeneity.  Figure \ref{fig:mod_plot} provides a visualization of the results of meta-regressions with our moderators.  Definitions of the subgroups for different moderator variables are included in Table S2, and more details of the regressions for other potential outcomes of interest are included in Table S7.

First, we found that the type of task significantly moderated human-AI synergy ($F(1,104) = 7.84$, two-tailed $p = 0.006$).  Among decision tasks---those in which participants decided between a finite set of options---the pooled effect size for human-AI synergy was significantly negative ($g = -0.27$, $t(104)=-3.20$, two-tailed $p=0.002$, 95\%CI $-0.44$ to $-0.10$), which indicates performance losses from combining humans and AI.  In contrast, among creation tasks---those in which participants created some sort of open response content---the pooled effect size for human-AI synergy was positive ($g = 0.19$, $t(104)=1.35$, two-tailed $p=0.180$, 95\%CI $-0.09$ to $0.48$), pointing to synergy between humans and AI. Even though the average performance gains for creation tasks were not significantly different from 0 (presumably because of the relatively small sample size of $n=34$), the difference between losses for decision tasks and gains for creation tasks was statistically significant. Relatedly, we found that the type of data involved in the task significantly moderated both human-AI synergy ($F(4, 101) = 15.24$, two-tailed $p = 0.000)$ and human augmentation ($F(4, 101) = 6.52$, two-tailed $p = 0.000)$). 

Second, we found that the performance of the human and AI relative to each other impacted both human-AI synergy ($F(1, 104) = 81.79$, two-tailed $p = 0.000$) and human augmentation ($F(1, 104) = 24.35$, two-tailed $p = 0.000$).  As shown in Figure \ref{fig:mod_plot}, when the human alone outperformed the AI alone, the combined human-AI system outperformed both alone with an average pooled effect size for human-AI synergy of $g = 0.46$ ($t(104) = 5.06$, two-tailed $p=0.000$, 95\%CI $0.28$ to $0.66$), a medium-sized effect~\cite{cohen2013statistical}). But when the AI alone outperformed the human alone, performance losses occurred in the combined system relative to the AI alone, with a negative effect size for human-AI synergy of $g = -0.54$ ($t(104) = -6.20$, two-tailed $p=0.000$, 95\%CI $-0.71$ to $-0.37$), a medium -sized effect~\cite{cohen2013statistical}).  Given the importance of this moderator, we fit separate meta-analytic models on the subset of results where (1) the AI performs better alone and (2) the human performs better alone, and we report the results for human-AI synergy and human augmentation in Table S4.

The performance of the human and AI relative to each other also affected the degree of human augmentation in the human-AI systems ($F(1, 104) = 24.35$, two-tailed $p = 0.000$).  When the AI outperformed the human alone, greater performance gains tended to occur in the human-AI systems relative to the human alone, and the pooled effect size for human augmentation was positive and medium to large in magnitude ($g = 0.74$, $t(104) = 13.50$, two-tailed $p=0.000$, 95\%CI $0.63$ to $0.85$) (see Figures S4-S7 for a more detailed visualization of this result for decision tasks).

Additionally, we found that the type of AI involved in the experiment ($F(2, 103) = 3.77$, two-tailed $p = 0.026$) and the year of publication ($F(3, 102) = 3.65$, two-tailed $p = 0.015$) moderated human-AI synergy, and the experimental design moderated human augmentation ($F(1, 104) = 4.90$, two-tailed $p = 0.029$). See Figure S8 for a more detailed visualization of the effect sizes by year of publication.

The remaining moderators we investigated were not statistically significant for human-AI synergy or human augmentation (explanation, confidence, participant type, division of labor).

\section{Discussion}
Systems that combine human intelligence and AI tools can address multiple issues of societal importance, from how we diagnose disease to how we design complex systems~\cite{shin2021ai, noti2022learning, chen2023understanding}. But some studies show how augmenting humans with AI can lead to better outcomes than humans or AI working alone~\cite{reverberi2022experimental, chen2023understanding, liu2021understanding}, while others show the opposite~\cite{bansal2021does, zhang2020effect, vaccaro2019the}. These seemingly disparate results raise two important questions: How effective is human-AI collaboration, in general? And under what circumstances does this collaboration lead to performance gains versus losses? Our study analyzes over three years of recent research to provide insights into both of these questions.

\textbf{Performance Losses from Human-AI Collaboration}: Regarding the first question, we found that, on average among recent experiments, human-AI systems did not exhibit synergy: the human-AI groups performed worse than either the human alone or the AI alone.  This result complements the qualitative literature reviews on human-AI collaboration~\cite{lai2023towards, sperrle2021survey, maadi2021review}, which highlight some of the surprising challenges that arise when integrating human and artificial intelligence.  For example, people often rely too much on AI systems ("overreliance"), using its suggestions as strong guidelines without seeking and processing more information~\cite{skitka1999does, buccinca2021to, lai2020why}.  Other times, however, humans rely too little on AI ("underreliance"), ignoring its suggestions because of adverse attitudes toward automation~\cite{zhang2020effect, buccinca2021to, vasconcelos2023explanations}.

Interestingly, we found that, among this same set of experiments, human augmentation did exist in the human-AI systems: the human-AI groups performed better than the humans working alone.  Thus, even though, on average, the human-AI combinations did not achieve synergy, the AI system did, on average, help humans perform better. This result can occur, of course, because by definition the baseline for human-AI synergy is more stringent than that for human augmentation.  It may also occur, however, because obtaining human-AI synergy requires different forms of human-AI interaction, or because the recent empirical studies were not appropriately designed to elicit human-AI synergy.

\textbf{Moderating Effect of Task Type:} With the large data set we collected, we also performed additional analyses of factors that influence the effectiveness of human-AI collaboration.  We found that the type of task significantly moderated synergy in human-AI systems with decision tasks associated with performance losses and creation tasks associated with performance gains. 

We hypothesize that this advantage for creation tasks occurs because even when creation tasks require the use of creativity, knowledge, or insight for which humans perform better, they often also involve substantial amounts of somewhat routine generation of additional content that AI can perform as well or better than humans. For instance, generating a good artistic image usually requires some creative inspiration about what the image should look like, but it also often requires a fair amount of more routine fleshing out of the details of the image. Similarly, generating many kinds of text documents often requires knowledge or insight that humans have and computers do not, but it also often requires filling in boilerplate or routine parts of the text as well. 

With most of the decision tasks studied in our sample, however, both the human and the AI system make a complete decision, with the humans usually making the final choice. Our results suggest that with these decision tasks, better results might have been obtained if the experimenters had designed processes in which the AI systems did only the parts of the task for which they were clearly better than humans. Only 3 of the 100+ experiments in our analysis explore such processes with a pre-determined delegation of separate sub-tasks to humans and AI.  With the 4 effect sizes from these 3 experiments, we found that, on average, human-AI synergy ($g = 0.22$, $t(104)=0.69$, two-tailed $p=0.494$, 95\% CI $-0.42$ to $0.87$) occurred, but the result was not statistically significant (see Section S2.6 for a more detailed discussion of these experiments).  

\textbf{Moderating Effect of Relative Human/AI Performance}: Interestingly, when the AI alone outperformed the human alone, substantial performance \textit{losses} occurred in the human-AI systems.  When the human outperformed the AI alone, however, performance \textit{gains} occurred in the human-AI systems.  This finding shows that the human-AI performance cannot be explained with a simple average of the human alone and AI alone.  In such a case, human-AI synergy could never exist~\cite{donahue2022human}.

Most (>95\%) of the human-AI systems in our data set involved humans making the final decisions after receiving input from the AI algorithms. In these cases, one potential explanation of our result is that, when the humans are better than the algorithms overall, they are also better at deciding in which cases to trust their own opinions and in which to rely more on the algorithm’s opinions. 

For example, \cite{cabrera2023improving} use an experimental design in which subjects in the human-AI condition saw a problem instance, an AI prediction for that instance, and in some cases, additional descriptions of the accuracy of the AI on this type of instance. The same experimental design, with the same task interface, participant pool, and accuracy of the AI system, was used for three separate tasks: fake hotel review detection, satellite image classification, and bird image classification.  For fake hotel review detection, the researchers found that the AI alone achieves an accuracy of 73\%, the human alone achieves an accuracy of 55\%, and the human-AI system achieves an accuracy of 69\%.  In this case, we hypothesize that, since the people are less accurate, in general, than the AI algorithms, they are also not good at deciding when to trust the algorithms and when to trust their own judgement, so their participation results in lower overall performance than for the AI algorithm alone.  

On the other hand, \cite{cabrera2023improving} find that, for bird image classification, the AI alone achieves an accuracy of 73\%, the human alone achieves an accuracy of 81\%, and the human-AI system achieves an accuracy of 90\%.  Here, the humans alone are more accurate than the AI algorithms alone, so we hypothesize that the humans are good at deciding when to trust their own judgements versus those of the algorithms, and, thus, the overall performance improves over either humans or AI alone.

\textbf{Surprisingly Insignificant Moderators:} We also investigated other moderators such as the presence of an explanation, the inclusion of the confidence of the AI output, and the type of participant evaluated.  These factors have received much attention in recent years~\cite{bansal2021does, chen2023understanding, liu2021understanding, rastogi2022deciding}.  Given our result that, on average across our 300+ effect sizes, they do not impact the effectiveness of human-AI collaboration, we think researchers may wish to de-emphasize this line of inquiry and instead shift focus to the significant and less researched moderators we identified: the baseline performance of the human and AI alone, the type of task they perform, and the division of labor between them. 

\textbf{Limitations:} Importantly, we want to highlight some general limitations of our meta-analytic approach to aid with the interpretation of our results.  First, our quantitative results apply to the subset of studies we collected through our systematic literature review. To evaluate human-AI synergy, we required that papers report the performance of (1) the human alone, (2) the AI alone, and (3) the human-AI system. We can, however, imagine tasks that a human and / or AI cannot perform alone but can when working with the other. Our analysis does not include such studies.  

Second, we calculate effect sizes that correspond to different quantitative measures such as task accuracy, error, and quality.  By computing Hedges' $g$, a unitless standardized effect size, we can describe important relations among these experiments in ways that make them comparable across different study designs with different outcome variables~\cite{hedges2008effect}.  The studies in our data set, though, come from different samples of people -- some look at doctors~\cite{jacobs2021machine, tschandl2020human, jussupow2021augmenting}, others at crowdworkers~\cite{bansal2021does, cabrera2023improving, lai2020why}, and still others at students~\cite{he2023interaction, liang2022adapting, boskemper2022measuring} –- and this variation can limit the comparability of the effect size to a degree~\cite{hedges2008effect}.  Additionally, the measurement error can also vary across experiments.  For example, some studies estimate overall accuracy based on the evaluation of as many as 500 distinct images~\cite{reverberi2022experimental} whereas others estimate it based on the evaluation of as few as 15 distinct ones~\cite{papenmeier2022its}.  As is typical for meta-analyses~\cite{borenstein2021introduction}, in our three-level model, we weight effect sizes as a function of their variance across participants, so we do not account for this other source of variation in measurement. 

Third, although we did not find evidence of publication biases, it remains possible that they exist, which would impact our literature base and, by extension, our meta-analytic results. However, we expect that if there were a publication bias operating here, it would be a bias to publish studies that showed significant gains from combining humans and AI. And since our overall results showed the opposite, it seems unlikely that they are a result of publication bias.

Fourth, our results only apply to the tasks, processes, and subject pools that researchers have chosen to study, and these configurations may not be representative of the ways human-AI systems are configured in practical uses of AI outside of the laboratory. In other words, even if there is not a publication bias in the studies we analyzed, there might be a research topic selection bias at work.

Fifth, the quality of our analysis depends on the quality of the studies we synthesized.  We tried to control for this issue by only including studies published in peer-reviewed publications, but the rigor of the studies may still vary in degree.  For example, studies used different attention check mechanisms and performance incentive structures, which can both affect the quality of responses and thus introduce another source of noise into our data.

Finally, we find a high level of heterogeneity among the effect sizes in our analysis.  The moderators we investigate account for some of this heterogeneity, but much remains unexplained.  We hypothesize that interaction effects exist between the variables we coded (e.g. explanation and type of AI), but we do not have enough studies to detect such effects.  There are also certainly potential moderators that we did not analyze.  For example, researchers mostly used their own experimental platforms and stimuli, which naturally introduce sources of variation between their studies.  As the human-AI collaboration literature develops, we hope future work can identify more factors that influence human-AI synergy and assess the interactions among them.

\textbf{A Roadmap for Future Work: Finding Human-AI Synergy:} Even though our main result suggests that -- on average -- combining humans and AI leads to performance losses, we do not think this means that combining humans and AI is a bad idea.  On the contrary, we think it just means that future work needs to focus more specifically on finding effective processes that integrate humans and AI. Our other results suggest promising ways to proceed.

\textit{Develop Generative AI for Creation Tasks.} In our broad sample of recent experiments, the vast majority (about 85\%) of the effect sizes were for decision-making tasks in which subjects chose among a predefined set of options. But in these cases we found that the average effect size for human-AI synergy was significantly negative. On the other hand, only about 10\% of the effect sizes researchers studied were for creation tasks---those that involved open-ended responses. And in these cases we found that the average effect size for human-AI synergy was positive and significantly greater than that for decision tasks. This result suggests that studying human-AI synergy for creation tasks---many of which can be done with generative AI---could be an especially fruitful area for research.

Much of the recent work on generative AI with human-subjects, however, tends to focus on attitudes towards the tool~\cite{wilcox2023ai, karinshak2023working}, interviews or think-alouds with participants~\cite{vimpari2023adapt, liu2023experiences, jo2023understanding}, or user experience instead of task performance~\cite{petridis2023anglekindling, jakesch2023co, mirowski2023co}.  Furthermore, the relatively little work that does evaluate human-AI collaboration according to quantitative performance metrics tends to only report the performance of the human alone and human-AI combination (not the AI alone)~\cite{noy2023experimental}.  This limitation makes evaluating human-AI synergy difficult as the AI alone may be able to perform the task at a higher quality and speed than the participants involved in the experiment, typically crowdworkers.  We thus need studies that further explore human-AI collaboration across diverse tasks while reporting the performance of the human alone, AI alone, and human-AI system. 

\textit{Develop Innovative Processes.} Additionally, as discussed in~\cite{donahue2022human}, human-AI synergy requires that humans be better at some parts of a task, AI be better at other parts of the task, and the system as a whole be good at appropriately allocating subtasks to whichever partner is best for that subtask. Sometimes that is done by letting the more capable partner decide how to do allocation of subtasks, and sometimes it is done by assigning different subtasks a priori to the most capable partner (see Section S2.6 of the SI for specific examples from experiments in our dataset). In general, to effectively use AI in practice, it may be just as important to design innovative processes for how to combine humans and AI as it is to design innovative technologies~\cite{baier2024your}. 

\textit{Develop More Robust Evaluation Metrics for Human-AI Systems.} Many of the experiments in our analysis evaluate performance according to a single measure of overall accuracy, but this measure corresponds to different things depending on the situation, and it omits other important criteria for human-AI systems.  For example, as one approaches the upper bound of performance, such as 100\% accuracy, the improvements necessary to increase performance usually become more difficult for both humans and AI systems.  In these cases, we may wish to consider a metric that applies a non-linear scaling to the overall classification accuracy and thus takes such considerations into account~\cite{campero2022test}.  

More importantly, there are many practical situations where good performance depends on multiple criteria. For instance, in many high-stakes settings such as radiology diagnoses and bail predictions, relatively rare errors may have extremely high financial or other costs. In these cases, even if AI can, on average, perform a task more accurately and less expensively than humans, it may still be desirable to include humans in the process if the humans are able to reduce the number of rare but very undesirable errors. One potential approach for situations like these is to create composite performance measures that incorporate the expected costs of various kinds of errors. The human augmentation measure described is also appropriate for these high-stakes settings. 

In general, we encourage researchers to develop, employ, and report more robust metrics that take into account factors such as task completion time, financial cost, and the practical implications of different types of errors.  These developments will help us better understand the significance of improvements in task performance as well as the effects of human-AI collaborations.

\textit{Develop Commensurability Criteria.} 
As researchers continue to study human-AI collaboration, we also urge the field to develop a set of commensurability criteria, which can facilitate more systematic comparisons across studies and help us track progress in finding areas of human-AI synergy.  These criteria could provide standardized guidelines for key study design elements such as:
\begin{enumerate}[nosep]
    \item Task Designs: Establish a set of benchmark tasks that involve human-AI systems.
    \item Quality Constraints: Specify acceptable quality thresholds or requirements that the human, AI, and human-AI system must meet.
    \item Incentive Schemes: Outline incentive structures (e.g. payment schemes, bonuses) used to motivate human participants.
    \item Process Types: Develop a taxonomy of interaction protocols, user interface designs, and task workflows for effective human-AI collaboration.
    \item Evaluation Metrics: Report the performance of the human, AI, and human-AI system according to well-defined performance metrics.
\end{enumerate}

To further promote commensurability and research synthesis, we encourage the field to establish a standardized and open reporting repository, specifically for human-AI collaboration experiments. This centralized database would host the studies' raw data, code, system outputs, interaction logs, and detailed documentation, adhering to the proposed reporting guidelines. As such, it would facilitate the replication, extension, and synthesis of research in the field. For example, by applying advanced machine learning techniques on such a dataset, we could develop predictive models to guide the design of human-AI systems optimized for specific constraints and contexts. Additionally, it would provide a means to track progress in finding greater areas of human-AI synergy.

In conclusion, our results demonstrate that human-AI systems often perform worse than humans alone or AI alone.  But our analysis also suggests promising directions for the future development of more effective human-AI systems.  We hope that this work will help guide progress in developing such systems and using them to solve some of our most important problems in business, science, and society. 


\section{Methods}
We conducted this meta-analysis in accord with the guidelines from Kitchenham 2004 on systematic reviews, and we follow the standards set forth by the Preferred Reporting Items for Systematic Reviews and Meta-Analyses (PRISMA)~\cite{moher2009preferred}.

\subsection{Literature Review}\label{lit_review}
\subsubsection{Eligibility Criteria}
Per our pre-registration (see \href{https://osf.io/wrq7c/?view_only=b9e1e86079c048b4bfb03bee6966e560}{here}), we applied the following criteria to select studies that fit our research questions.  First, the paper needed to present an original experiment that evaluates some instance in which a human and an AI system work together to perform a task.  Second, it needed to report the performance of (1) the human alone, (2) the AI alone, and (3) the human-AI system according to some quantitative measure(s).  As such, we excluded studies that reported the performance of the human alone but not the AI alone, and likewise we excluded studies that reported the performance of the AI alone but not the human alone.  Following this stipulation, we also excluded purely meta-analyses and literature reviews, theoretical work, qualitative analyses, commentaries, opinions, and simulations.  Third, we required the paper to include the experimental design, the number of participants in each condition, and the standard deviation of the outcome in each condition, or enough information to calculate it from other quantities.  Finally, we required the paper to be written in English.

\subsubsection{Search Strategy}
Given the interdisciplinary nature of human-AI interaction studies, we performed this search in multiple databases covering conferences and journals in the computer sciences, information sciences, and social sciences, as well as other fields.  Through consultation with a library specialist in these fields, we decided to target the Association for Computing Machinery Digital Library (ACM DL), Association for Information Systems eLibrary (AISeL), and the Web of Science Core Collection (WoS) for our review.  To focus on current forms of artificial intelligence, we limited the search to studies published between January 1, 2020 and June 30, 2023.  

To develop the search string, we began by distilling the facets of studies that evaluated the performance of a human-AI system. We required the following: (1) a human component, (2) a computer component, (3) a collaboration component, and (4) an experiment component.  Given the multidisciplinary nature of our research question, papers published in different venues tended to refer to these components under various names~\cite{bansal2019updates, boskemper2022measuring, groh2022deepfake, tejeda2022ai, bo2021toward}.  For broad coverage, we compiled a list of synonyms and abbreviations for each component and then combined these terms with Boolean operations, resulting in the following search string:

\begin{compactenum}
    \item \textbf{Human:} human OR expert OR participant OR humans OR experts OR participants
    \item \textbf{AI:} AI OR "artificial intelligence" OR ML OR "machine learning" OR "deep learning"
    \item \textbf{Collaboration:} collaborate OR assist OR aid OR interact OR help
    \item \textbf{Experiment:} "experiment" OR "experiments" OR "user study" OR "user studies" OR "crowdsourced study" OR "crowdsourced studies" OR "laboratory study" OR "laboratory studies"
\end{compactenum}

Search term: (1) AND (2) AND (3) AND (4) in the abstract.  See Table S1 for the exact syntax for each database.

To further ensure comprehensive coverage, we also conducted a forward and backwards search on all studies we found that meet our inclusion criteria. 

\subsubsection{Data Collection and Coding}
We conducted the search in each of these databases in July 2023.  To calculate our primary outcome of interest -- the effect of combining human and artificial intelligence on task performance -- we recorded the averages and standard deviations of the task performance of the human alone, the AI alone, and the human and AI working with each other, as well as the number of subjects in each of these conditions (see S1.3 for details).  

Many authors reported all of these values directly in the text of the paper.  A notable number, however, reported them indirectly by providing 95\% confidence intervals or standard errors instead of the raw standard deviations.  For these, we calculated the standard deviations using the appropriate formulas~\cite{higgins2019choosing}.  Additionally, multiple papers did not provide the exact numbers needed for such formulas, but the authors made the raw data of their study publicly accessible.  In these cases, we download the datasets and computed the averages and standard deviations using Python or R.  If relevant data were only presented in the plots of a paper, we contacted the corresponding author to ask for the numeric values.  If the authors did not respond, we used a Web Plot Digitizer~\cite{rohatgi2020webplotdigitizer} to convert plotted values into numerical values.  For papers that conducted an experiment that met our inclusion criteria but did not report all values need to calculate the effect size, we also contacted the corresponding author directly to ask for the necessary information.  If the author did not respond, we could not compute the effect size for the study and did not include it in our analysis. 

We considered and coded for multiple potential moderators of human-AI performance, namely: (1) publication date (2) pre-registration status (3) experimental design (4) data type (5) task type (6) task output (7) AI type (8) AI explanation (9) AI confidence (10) participant type and (11) performance metric.  See Table S2 for a description of each of these moderator variables.  The additional information we recorded served descriptive and exploratory purposes. 

Many papers conducted multiple experiments, contained multiple treatments, or evaluated performance according to multiple measures.  In such cases, we assigned a unique experiment identification number, treatment identification number, and measure identification number to the effect sizes from the paper.  Note that we defined experiments based on samples of different sets of participants.

\subsection{Data Analysis}\label{data_analysis}
\subsubsection{Calculation of Effect Size}
We computed Hedges' $g$ to measure the effect of combining human and artificial intelligence on task performance~\cite{hedges1981distribution}.  For strong synergy, Hedges' $g$ represents the standardized mean difference between the performance of the human-AI system performance and the baseline, which can be the human alone or AI alone, whichever one performs better, on average.  For human augmentation, Hedges' $g$ represents the standardized mean difference between the performance of the human-AI system and the baseline of the human alone.  

We chose Hedges' $g$ as our measure of effect size because it is unitless, so allows us to compare human-AI performance across different metrics, and it corrects for upward bias in the raw standardized mean difference (Cohen’s $d$)~\cite{hedges1981distribution}.  See Section S1.4 in the SI for more details about this calculation.

\subsubsection{Meta-Analytic Model}
The experiments from the papers we analyzed varied considerably.  For example, they evaluated different tasks, they recruited participants from different backgrounds, and they employed different experimental designs.  Since we expected substantial between-experiment heterogeneity in the true effect sizes, for our analysis we used a random-effects model that accounts for variance in effect sizes that comes from both sampling error and "true" variability across experiments~\cite{hedges2014statistical}.  

Additionally, some of the experiments we considered generated multiple dependent effect sizes: they could involve multiple treatment groups, and they could assess performance according to more than one measure, for example.  The more commonly used meta-analytic models assume independence of effect sizes, which makes them unsuitable for our analysis~\cite{van2015meta}.  We thus adopted a three-level meta-analytic model in which effect sizes are nested within experiments, so we explicitly took the dependencies in our data into account in the model~\cite{van2015meta, cheung2014modeling}.  Furthermore, we used robust variance estimate (RVE) methods to compute consistent estimates of the variance of our effect sizes estimates and, relatedly, standard errors, confidence intervals, and statistical tests, which account for the dependency of sampling errors from overlapping samples that occurred in experiments that compared multiple treatment groups to a single control group~\cite{hedges2010robust}.  When evaluating the significance of our result, we applied the Knapp and Hartung adjustment and computed a test statistic, standard error, $p$-value, and confidence interval based on the $t$-distribution with  degrees of freedom, where  denotes the number of effect size clusters (i.e. number of experiments) and  denotes the number of coefficients in the model.  To perform our moderator analyses, we conducted separate meta-regressions for each of our moderator variables.

To interpret the degree of heterogeneity in our effect sizes, we calculated the popular $I^2$ statistic following~\cite{higgins2002quantifying}, which quantifies the percentage of variation in effect sizes that is not from sampling error.  Furthermore, to distinguish the sources of this heterogeneity, we also calculated multilevel versions of the statistic, following~\cite{cheung2014modeling}.

\subsection{Bias Tests}
In the context of our meta-analysis, publication bias may occur if researchers publish experiments that show evidence of significant human-AI synergy more frequently than those that do not.  Such actions would affect the data we collected and distort our findings.  To evaluate this risk, we adopted multiple diagnostic procedures outlined by Viechtbauer and Cheung~\cite{viechtbauer2010outlier}.  First, we created funnel plots that graph the observed effect sizes on the $x$-axis and the corresponding standard errors on the $y$-axis~\cite{sterne2001funnel}.  In the absence of publication bias, we expect the points to fall roughly symmetrically around the y-axis.  We enhanced these plots with colors indicating the significance level of each effect size to help distinguish publication bias from other causes of asymmetry~\cite{peters2008contour}.  A lack of effect sizes in regions of statistical non-significance points to a greater risk of publication bias.  We visually inspected and performed Egger’s regression test~\cite{egger1997bias} as well as the rank correlation test~\cite{begg1994operating} to evaluate the results in the funnel plot.  These tests examine the correlation between the observed effect sizes and their associated sampling variances.  A high correlation indicates asymmetry in the funnel plot, which may stem from publication bias.

Figure S9 displays the funnel plot of the included effect sizes, and we do not observe significant asymmetry or regions of missing data in the plot for our primary outcome, human-AI synergy.  Egger’s regression did not indicate evidence of publication bias in the sample ($\beta=-0.67$, $t(104)=-0.78$, two-tailed $p=0.438$, 95\% CI $-2.39$ to $1.04$), nor did the rank correlation test ($\tau=0.05$,two-tailed $p=0.121$).  Taken as a whole, these tests suggest that our results for human-AI synergy are robust to potential publication bias.

Importantly, however, we do find potential evidence of publication bias in favor of studies that report results in which the human-AI system outperforms the human alone (human augmentation).  In this case, Egger’s regression does point to publication bias in the sample ($\beta=1.96$, $t(104)=3.24$, two-tailed $p=0.002$, 95\% CI $0.76$ to $3.16$), as does the rank correlation test ($\tau=0.19$,two-tailed $p=0.000$). Note that we did not try to correct for potential publication bias to preserve the integrity of the original data and maintain transparency in our reporting.  Many proposed adjustment methods can also lead to overcorrection and distort results~\cite{rothstein2005publication}.

The discrepancy between the symmetry in the funnel plot for human-AI synergy versus asymmetry in the funnel plot for human augmentation may reflect how many researchers and journals implicitly assume an interest in human augmentation, comparing the human-AI system to the human alone.

\subsection{Sensitivity Analysis}
For our primary analysis, we developed a three-level meta-analytic model that accounted for variance in the observed effect sizes (first level), variance between effect sizes from the same experiment (second level), and variance between experiments (third level).  We then calculated cluster-robust standard errors, confidence intervals, and statistical tests for our effect size estimates in which we defined clusters at the level of the experiment.  This model accounts for the dependency in effect sizes that result from evaluating more than one treatment against a common control group and assessing performance according to more than one measure.  It does, however, consider the experiments in our analysis as independent from each other, even if they come from the same paper.  We find this assumption plausible because the experiments recruited different sets of participants and entailed different tasks or interventions.  

As a robustness check, though, we performed a sensitivity re-analysis in which we clustered at the paper level instead of the experiment level.  This multilevel model accounts for variance in the observed effect sizes (first level), variance between effect sizes from the same paper (second level), and variance between papers (third level).  We still calculated cluster-robust standard errors, confidence intervals, and statistical tests for our effect size estimates in which we defined clusters at the level of the experiment because the participant samples only overlapped on the level of the experiment.  Using this approach, we found a comparable overall effect size for human-AI synergy ($g = -0.22$, $t(67)=-2.46$, two-tailed $p = 0.017$, 95\% CI $-0.41$ to $-0.04$) and for human augmentation ($g = 0.65$, $t(69)=9.96$, two-tailed $p = 0.000$, 95\% CI $0.52$ to $0.78$).  

We also evaluated the robustness of our results to outlying and influential data points.  To detect such data, we computed the residuals and Cook’s distance for each effect size.  We considered residual values greater or less than three standard deviations from the mean as outliers, and following~\cite{altman2016analyzing}, we considered values greater than $4/N$ as high influence, where $N$ is the number of data points in our analysis.  Using this approach, we identified 11 outliers for human-AI synergy and 9 outliers for human augmentation.  We performed a sensitivity re-analysis on a dataset excluding these effect sizes, which resulted in similar effect size for human-AI synergy ($g = -0.25$, $t(104)=-3.45$, two-tailed $p = 0.001$, 95\% CI $[-0.39, -0.11]$) and human augmentation ($g = 0.60$, $t(104)=12.60$, two-tailed $p = 0.000$, 95\% CI $[0.50, 0.69]$).   

Additionally, we conducted leave-one-out analyses, in which we performed a series of sensitivity re-analysis on the different subsets of effect sizes obtained by leaving out one effect size in our original dataset.  We also conducted leave-one-out analyses at the experiment and publication levels.  These tests show how each effect size, experiment, and publication affect our overall estimate of the effect of human-AI collaboration on task performance.  Summary effect sizes for human-AI synergy ranged from $-0.28$ to $-0.19$ with two-tailed $p < 0.05$ ($0.000$ to $0.019$) in all cases, indicating the robustness of our results to any single effect size, experiment, or paper; likewise, summary effect sizes for human augmentation ranged from $0.61$ to $0.66$ with two-tailed $p < 0.05$ ($0.000$ to $0.000$) in all cases, indicating the robustness of our results to any single effect size, experiment, or paper.


Lastly, we conducted a sensitivity re-analysis on a dataset that omits the effect sizes we estimated, either using WebPlotDigitizer or the information provided by the authors in their paper, and again we found almost identical results for human-AI synergy ($g = -0.21$, $t(98)=-2.59$, two-tailed $p = 0.011$, 95\% CI $[-0.36, -0.05]$) and human augmentation ($g = 0.64$, $t(98)=11.73$, two-tailed $p = 0.000$, 95\% CI $[0.53, 0.75]$).

We performed all quantitative analysis with the R statistical programming language, and we primarily relied on the package metafor~\cite{viechtbauer2010conducting}. 

\section{Data Availability}
We compiled the data used in this analysis based on the studies identified in our systematic literature review. We make the data we collected available via the project's Open Science Framework repository (\url{https://osf.io/wrq7c/?view_only=b9e1e86079c048b4bfb03bee6966e560}). In our systematic review, we searched the following databases: the ACM Digital Library (ACM DL) (\url{https://dl.acm.org/}), Web of Science (\url{https://clarivate.com/webofsciencegroup/solutions/web-of-science/}), and Association for Information Systems eLibrary
(AISeL)(\url{https://aisnet.org/page/AISeLibrary}).

\section{Code Availability}
We share the code used to conduct our analysis via the Open Science Framework repository (\url{https://osf.io/wrq7c/?view_only=b9e1e86079c048b4bfb03bee6966e560}).

\section{Acknowledgements}
MV was funded by the Accenture Technology Convergence Fellowship. Additional funding to TM was provided by the Toyota Research Institute, by the MIT Quest for Intelligence, and by the National Research Foundation, Prime Minister’s Office, Singapore under its Campus for Research Excellence and Technological Enterprise (CREATE) program.  The funding organizations had no role in study design, data collection and analysis, decision to publish or preparation of the manuscript. We thank Jessika Kim and Garrett Campagna for their help collecting the data for this project, and we thank Dean Eckles, Mohammed Alsobay, Robin Na, and Emily Hu for their helpful feedback.

\section{Author Contribution Statement}
MV conceived the study idea with feedback from AA and TM. MV collected the data and performed the statistical analysis, with feedback from AA and TM.  MV, AA, and TM wrote the manuscript.

\section{Competing Interest Statement}
The authors declare no competing interests.

\section{Figures}

\printbibliography[title={References}]

\begin{refsection}
\newpage

\setcounter{equation}{0}
\setcounter{figure}{0}
\setcounter{table}{0}
\setcounter{section}{0}
\setcounter{page}{1}
\renewcommand{\theequation}{S\arabic{equation}}
\renewcommand{\thefigure}{S\arabic{figure}}
\renewcommand{\thetable}{S\arabic{table}}
\renewcommand{\thesection}{S\arabic{section}}

\fancypagestyle{plain}{%
  \fancyhf{} 
  \renewcommand{\headrulewidth}{0pt} 
  \renewcommand{\footrulewidth}{0pt} 
  \fancyfoot[C]{S\thepage} 
}
\pagestyle{plain}

{\LARGE \textbf{Supplementary Information}}

\tableofcontents
\clearpage

\section{Supplementary Methods}
\subsection{Types of Outcomes in Human-AI Systems}\label{sec:synergy_def}
Let $H$, $AI$, and $HAI$ represent the performance of the human alone, AI alone, and human-AI combination, respectively. 

\begin{definition}[Human-AI Synergy]
The human-AI group outperforms both the human and the AI alone.
$$HAI > \max(H, AI)$$
\end{definition}

\begin{definition}[Human Augmentation]
The human-AI group outperforms the human alone.
$$HAI > H$$
\end{definition}

\begin{definition}[AI Augmentation]
The human-AI group outperforms the AI alone.
$$HAI > AI$$
\end{definition}

\begin{definition}[Negative Synergy]
The human-AI group outperforms neither the human or AI alone.
$$HAI < \min(H, AI)$$
\end{definition}

\subsection{Systematic Literature Review}

\begin{table}[H]
\begin{tabular}{p{.250\textwidth}p{.75\textwidth}}
    \toprule
\textbf{Database} & \textbf{Search String} \\ 
\midrule
ACM Digital Library (ACM DL) & [[Abstract: human] OR [Abstract: expert] OR [Abstract: participant] OR [Abstract: humans] OR [Abstract: experts] OR [Abstract: participants]] AND [[Abstract: ai] OR [Abstract: "artificial intelligence"] OR [Abstract: ml] OR [Abstract: "machine learning"] OR [Abstract: "deep learning"]] AND [[Abstract: collaborate] OR [Abstract: assist] OR [Abstract: aid] OR [Abstract: interact] OR [Abstract: help]] AND [[Abstract: "experiment"] OR [Abstract: "experiments"] OR [Abstract: "user study"] OR [Abstract: "user studies"] OR [Abstract: "crowdsourced study"] OR [Abstract: "crowdsourced studies"] OR [Abstract: "laboratory study"] OR [Abstract: "laboratory studies"]] AND [E-Publication Date: (01/01/2020 TO 06/30/2023)] \\ 
Web of Science Core Collection (WoS) & (((AB=(human OR expert OR participant OR humans OR experts OR participants)) AND AB=(AI OR "artificial intelligence" OR ML OR "machine learning" OR "deep learning")) AND AB=(collaborate OR assist OR aid OR interact OR help)) AND AB=("experiment" OR "experiments" OR "user study" OR "user studies" OR "crowdsourced study" OR "crowdsourced studies" OR "laboratory study" OR "laboratory studies") Index Date 2020-01-01 to 2023-06-30 \\  
Association for Information Systems eLibrary (AISeL) & abstract:( human OR expert OR participant OR humans OR experts OR participants ) AND abstract:( AI OR "artificial intelligence" OR ML OR "machine learning" OR "deep learning" ) AND abstract:( collaborate OR assist OR aid OR interact OR help ) AND abstract:( "experiment" OR "experiments" OR "user study" OR "user studies" OR "crowdsourced study" OR "crowdsourced studies" OR "laboratory study" OR "laboratory studies" ) Date Range = 
 (01/01/2020 TO 06/30/2023)\\  
  \bottomrule
\end{tabular}
    \caption{Syntax of the search strings for the literature review.}
    \label{tab:search_strings}
\end{table}

\begin{figure}[H]
    \centering
    \includegraphics[width=\textwidth]{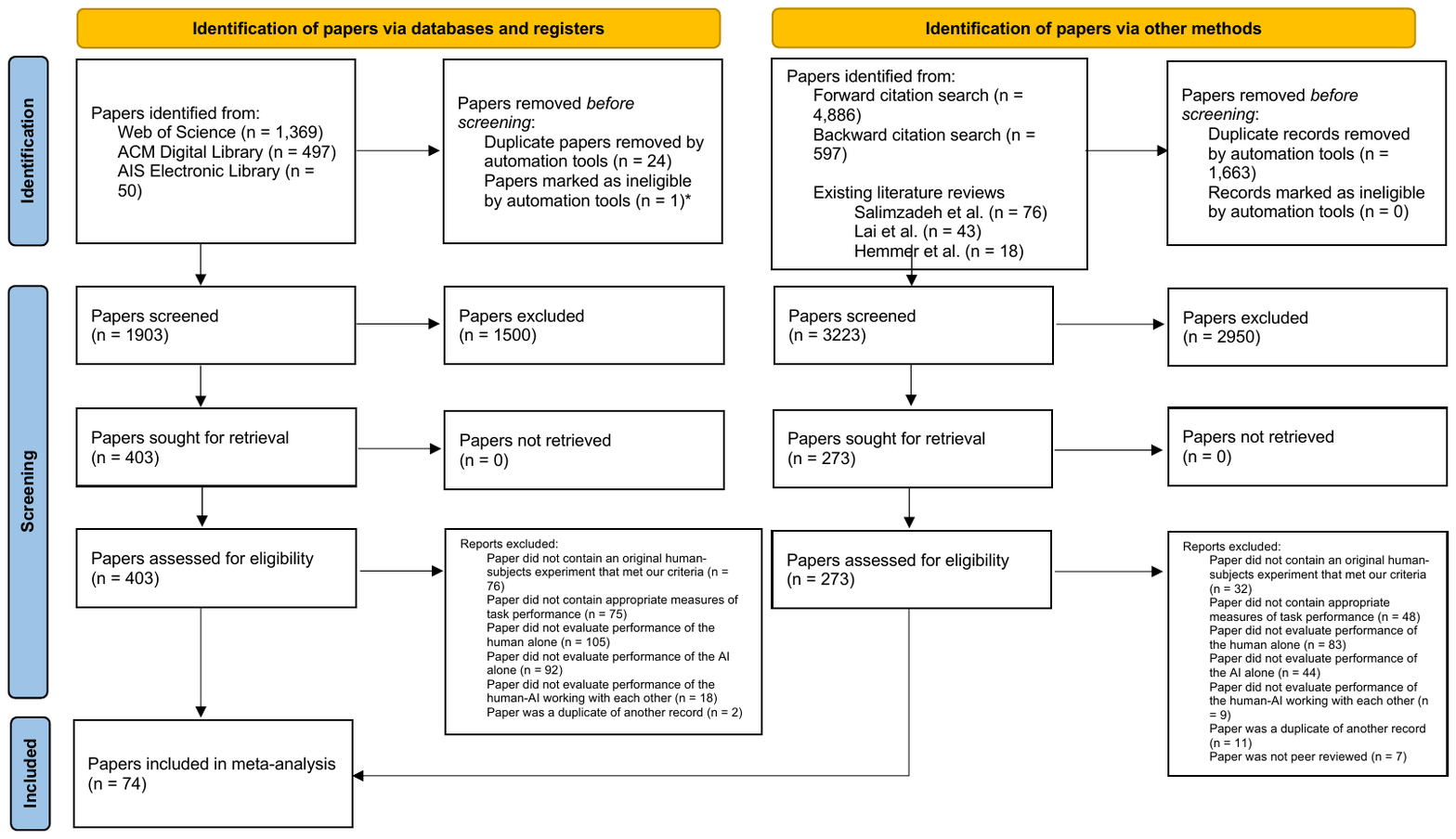}
    \caption{PRISMA flow diagram for the literature review and study inclusion process. $^*$Article retracted from journal. Adapted from~\cite{page2021prisma}.}
    \label{fig:prisma}
\end{figure}

\subsection{Data Collection and Coding}\label{data_calc}
We conducted the search in each of these databases in July 2023.  We downloaded all search results as .ris files and imported them into Zotero to check for duplicates.  Then, we exported the de-duplicated records into an Excel spreadsheet for screening.  We extracted the records that passed our initial abstract screen into a separate sheet in the Excel file.  Next, we evaluated if these records met our inclusion criteria by closely reading the paper.  If they did, we put them into a separate Excel file for data extraction.

To calculate our primary outcome of interest -- the effect of combining human and artificial intelligence on task performance -- we recorded the averages and standard deviations of the task performance of the human alone, the AI alone, and the human and AI working with each other, as well as the number of subjects in each of these conditions.  For example, in an experiment that involves $N$ participants completing $M$ trials (i.e., classifying $M$ images) and that evaluates the task performance according to some metric $x$, we extracted the following values:

$$\mu_P = \frac{1}{N} \sum_{i=1}^N x_{P,i}$$

$$\sigma_P = \sqrt{\frac{\sum_{i=1}^N (x_{P,i} - \bar{x_{P}} }{N}}$$

Where $P$ refers to the task performance of the human alone, the AI alone, or the human and AI working with each other:

$$P \in \{H, AI, HAI\}$$

And $x_{P,i}$ represents the task performance for participant $i$, defined as the average performance across the $M$ trials:

$$x_{P,i} = \frac{1}{M} \sum_{i=1}^M x_{P,i,j}$$

As these formulas make clear, we selected individual humans for our primary unit of analysis as we designed our study to capture patterns in human behavior, particularly how they work with AI tools.  

Many authors reported all of these values directly in the text of the paper.  A notable number, however, reported them indirectly by providing 95\% confidence intervals or standard errors instead of the raw standard deviations.  For these, we calculated the standard deviations using the appropriate formulas \cite{higgins2019choosing}.  

Additionally, multiple papers did not provide the exact numbers needed for such formulas, but the authors made the raw data of their study publicly accessible.  In these cases, we download the datasets and computed the averages and standard deviations using Python or R.  If relevant data were only presented in the plots of a paper, we contacted the corresponding author to ask for the numeric values.  If the authors did not respond, we used a Web Plot Digitizer \cite{rohatgi2020webplotdigitizer} to convert plotted values into numerical values.  

For papers that conducted an experiment that met our inclusion criteria but did not report all values need to calculate the effect size, we contacted the corresponding author directly to ask for the necessary information.  If the author did not respond, we could not compute the effect size for the study and did not include it in our analysis. 

We also considered and coded for multiple potential moderators of human-AI performance, as described in Table \ref{tab:mod_desc}. 

Many papers conducted multiple experiments, contained multiple treatments, or evaluated performance according to multiple measures.  In such cases, we assigned a unique experiment identification number, treatment identification number, and measure identification number to the effect sizes from the paper.  Note that we defined experiments based on samples of different sets of participants.

\subsection{Calculation of Effect Size}\label{calc_effect}
For each evaluation metric, we noted whether higher or lower values indicate better or worse performance on the task.  For example, increases in the accuracy of a task indicate performances improvements, while increases the time to complete a task indicate performance losses.  In the latter cases, we multiplied the performances of the human, AI, and human-AI combination by negative one for consistent interpretation of effect sizes consistent across our dataset. 

As such, when Hedges' $g$ equals zero, the average performance of the human-AI system and the baseline are the same, which indicates no gains from human-AI collaboration.  A positive value of Hedges' $g$ means that the average performance of the human-AI system exceeds that of the baseline, which points to synergy between humans and AI.  In contrast, a negative value of Hedges' $g$ means that the average performance of the human-AI system falls below that of the baseline, which points to performance losses from human-AI collaboration.  Larger absolute values of Hedges' $g$ indicate larger effects of human-AI collaboration on task performance.  According to conventional interpretations, values of Hedges' $g$ around 0.2 correspond to a small effect, values around 0.5 a medium effect, and values around 0.8 a large effect \cite{cohen2013statistical}.

\begin{table}[H]
\begin{tabular}{p{.25\textwidth}p{.75\textwidth}}
    \toprule
\textbf{Moderator} & \textbf{Description} \\ 
\midrule
*Publication Date & Year the paper was published (2020, 2021, 2022, 2023) \\ 
*Experimental Design & Type of design of the experiment 
\begin{enumerate}[noitemsep,topsep=0pt]
    \item Between-subjects
    \item Within-subjects
    \item Mixed, between-subjects (factorial design where the human vs. human-AI and AI vs. human-AI comparisons were the between-subjects factor)
    \item Mixed, within-subjects (factorial design where the human vs. human-AI and AI vs. human-AI comparisons were the within-subjects factor)
\end{enumerate} 
\textit{Note that in our moderator analysis, we combined ``Between-subjects'' and ``Mixed, between-subjects'' (and likewise ``Within-subjects'' and ``Mixed, within-subjects'')}.\\ 
*Data Type & Type of data involved in the experimental task (Binary, Categorical, Image, Numeric, Text, Video)
\newline
\textit{Note the if there were multiple types of data involved in the experiment (i.e. an image and a piece of text), we recorded them both (i.e. ``Image, Text'')}.
\\ 
*Task Type & Type of task evaluated 
\begin{enumerate}[noitemsep,topsep=0pt]
    \item Creation: Does it involve some type of open response?
    \item Decision: Does it involve deciding between a set of options or providing a numerical value?
\end{enumerate}\\ 
*AI Type & Type of AI involved in the experiment 
\begin{enumerate}[noitemsep,topsep=0pt]
    \item Shallow model
    \item Deep learning model
    \item Wizard of Oz model
\end{enumerate} 
\textit{Note that, here, we follow Lai et al.'s classification scheme \cite{lai2023towards}}.\\ 
*AI Explanation & Whether or not an explanation of the AI model’s output was communicated to participants (Yes, No) \\ 
*Participant Type & Type of participant involved in the experiment 
\begin{enumerate}[noitemsep,topsep=0pt]
    \item Crowdworker / non-crowdworker
    \item Expert / non-expert
\end{enumerate}\\ 
*Division of Labor & Was there some pre-determined division of labor between the human and AI? \\ 
Task Output & Type of output from the task (Binary, Categorical, Image, Numeric, Text, Video) \\ 
AI Confidence & Whether or not the confidence of the AI model was communicated to participants (Yes, No) \\  
Who Performed Better? & Who performed better alone on the task (Human, AI) \\
  \bottomrule
\end{tabular}
    \caption{Descriptions of the moderator variables. *Indicates a pre-registered moderator variable.}
    \label{tab:mod_desc}
\end{table}

\section{Supplementary Results}

\subsection{Descriptive Statistics}
\begin{figure}[H]
\centering
\begin{subfigure}[t]{.32\textwidth}
    \centering
    \includegraphics[width=\linewidth]{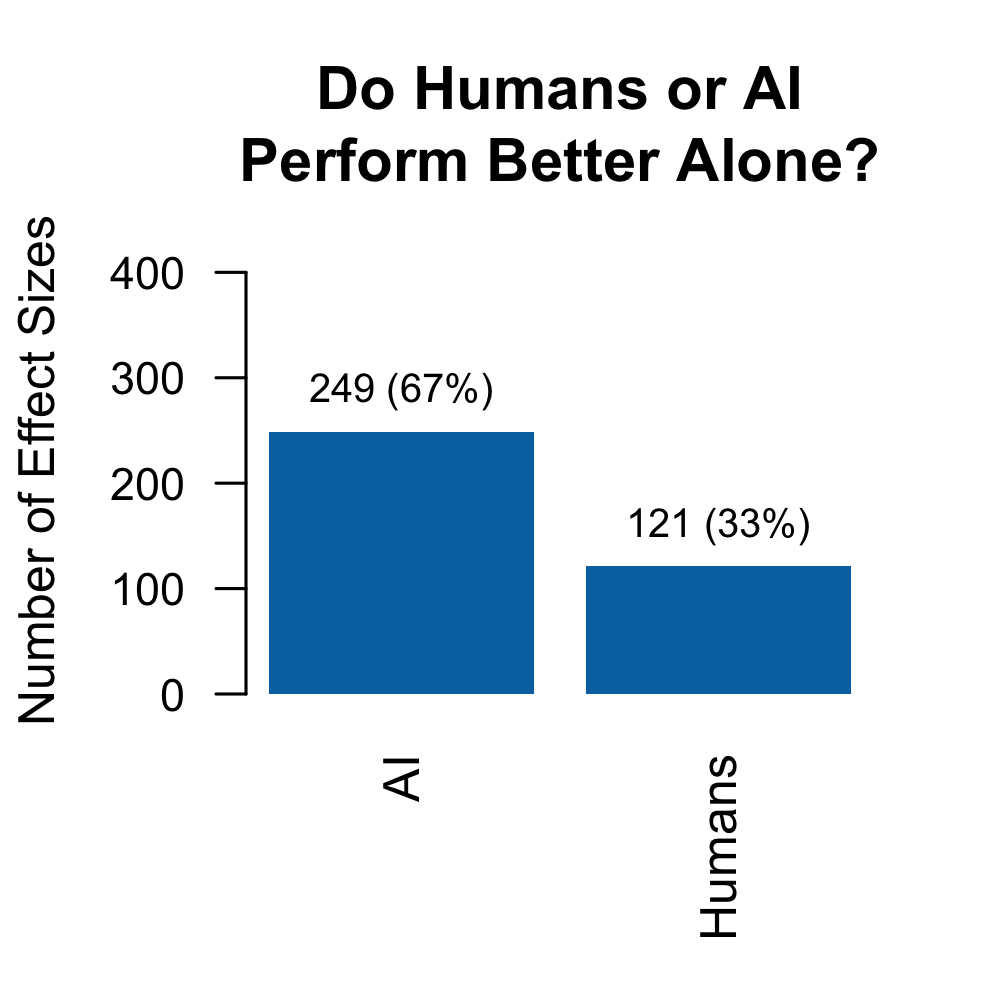}
\end{subfigure}
\hfill
\begin{subfigure}[t]{.32\textwidth}
    \centering
    \includegraphics[width=\linewidth]{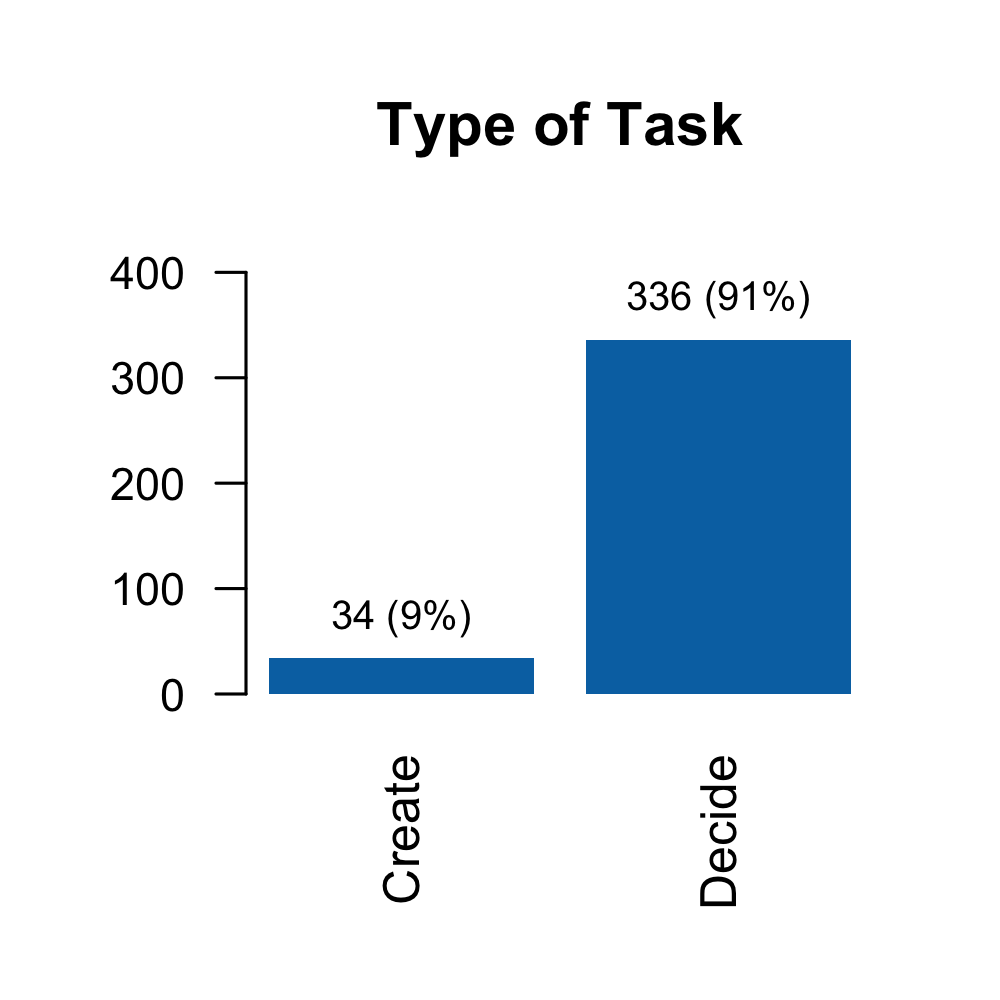}
\end{subfigure}
\hfill
\begin{subfigure}[t]{.32\textwidth}
    \centering
    \includegraphics[width=\linewidth]{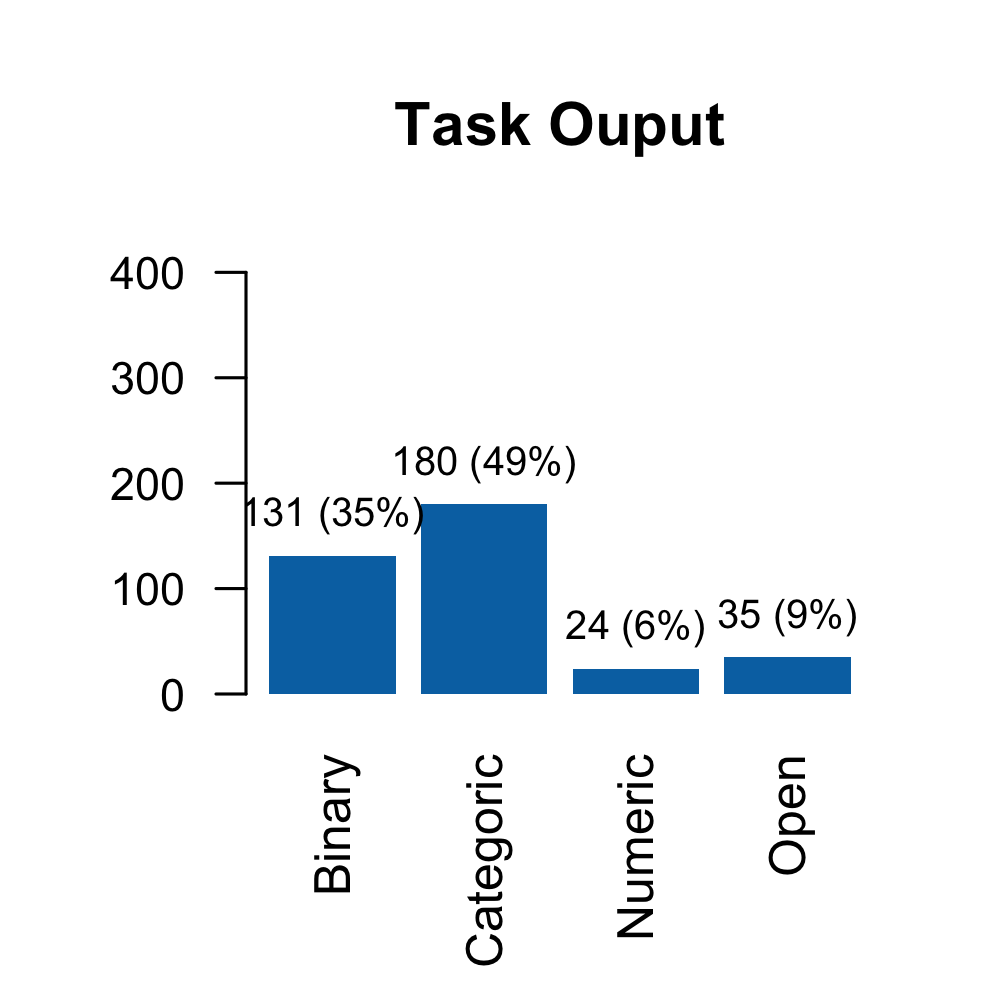}
\end{subfigure}
\hfill
\begin{subfigure}[t]{.32\textwidth}
    \centering
    \includegraphics[width=\linewidth]{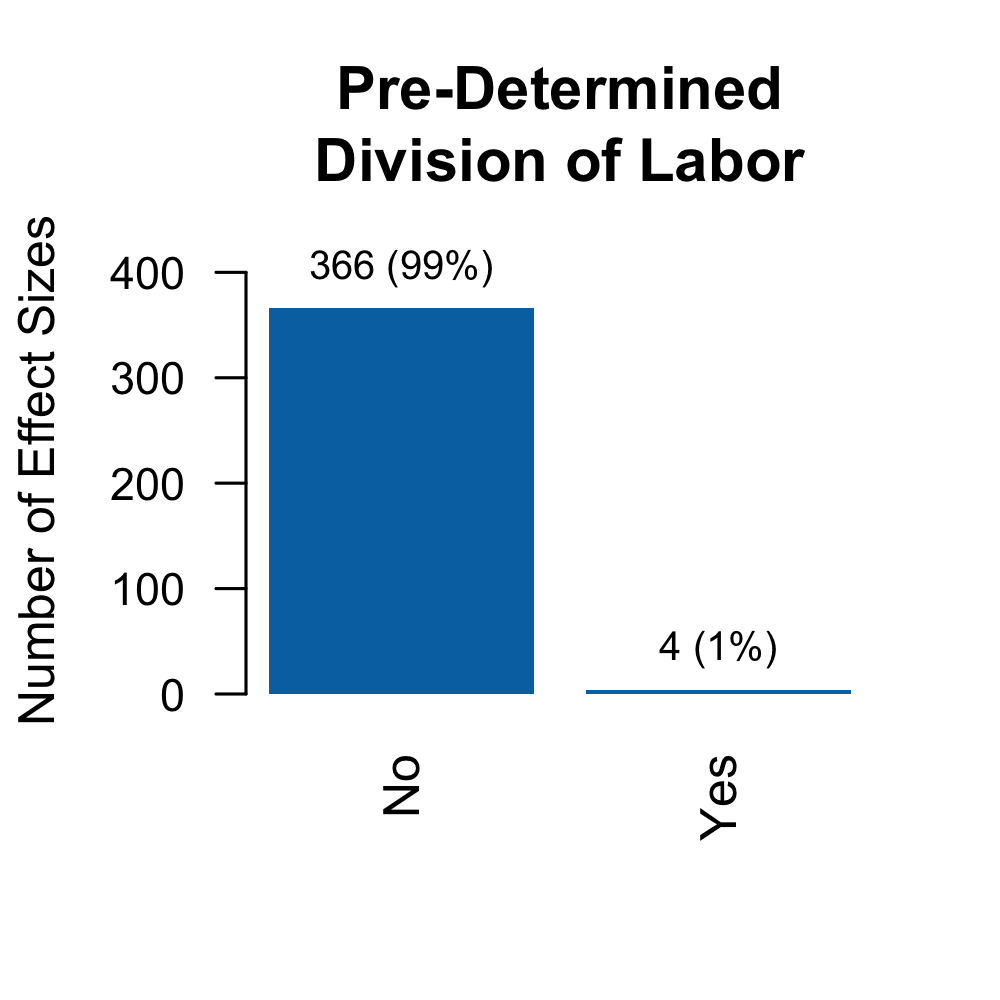}
\end{subfigure}
\centering
\begin{subfigure}[t]{.32\textwidth}
    \centering
    \includegraphics[width=\linewidth]{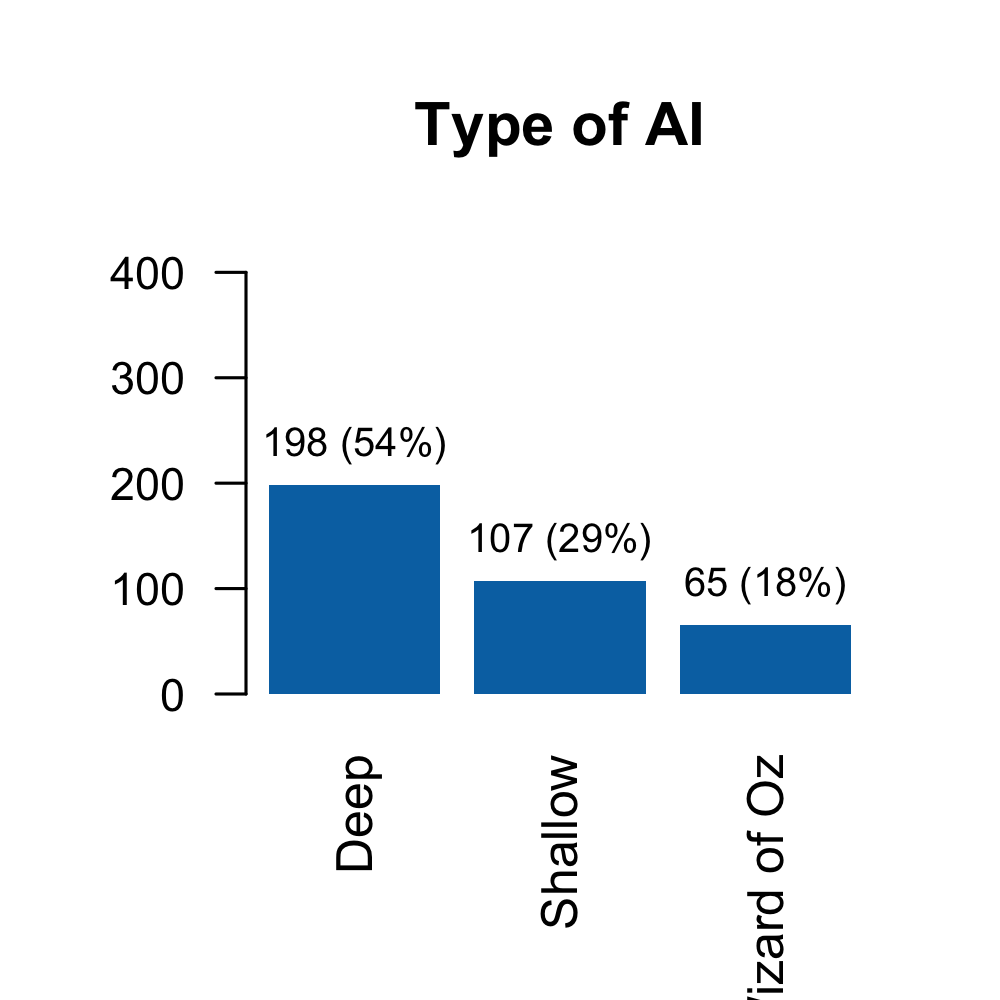}
\end{subfigure}
\hfill
\begin{subfigure}[t]{.32\textwidth}
    \centering
    \includegraphics[width=\linewidth]{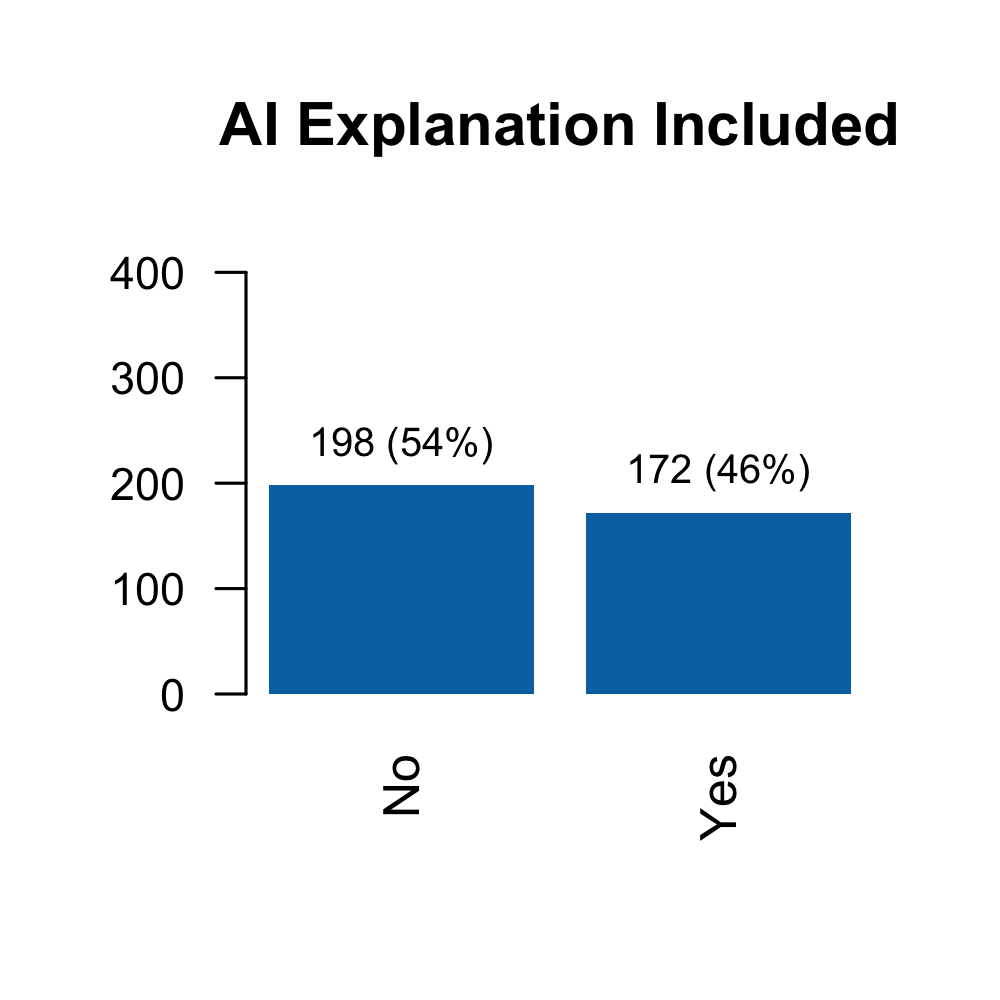}
\end{subfigure}
\hfill
\begin{subfigure}[t]{.32\textwidth}
    \centering
    \includegraphics[width=\linewidth]{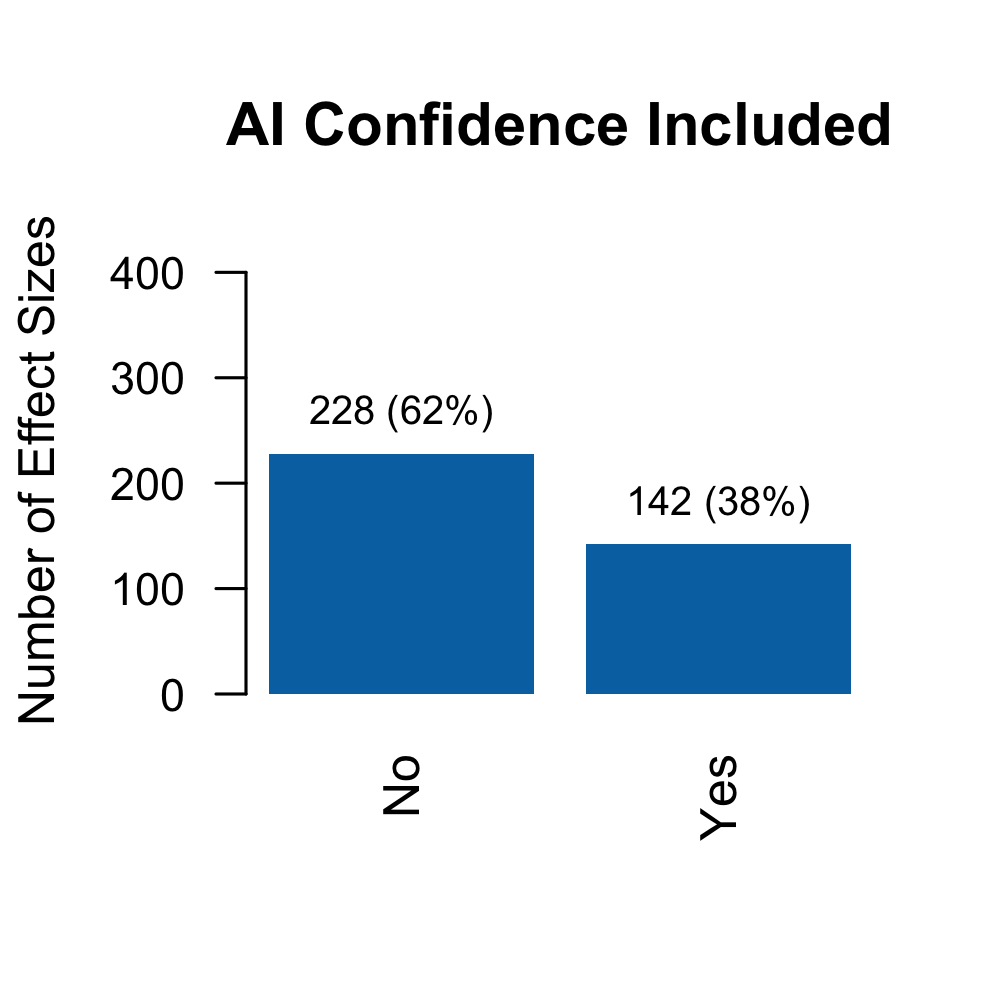}
\end{subfigure}
\centering
\begin{subfigure}[t]{.32\textwidth}
    \centering
    \includegraphics[width=\linewidth]{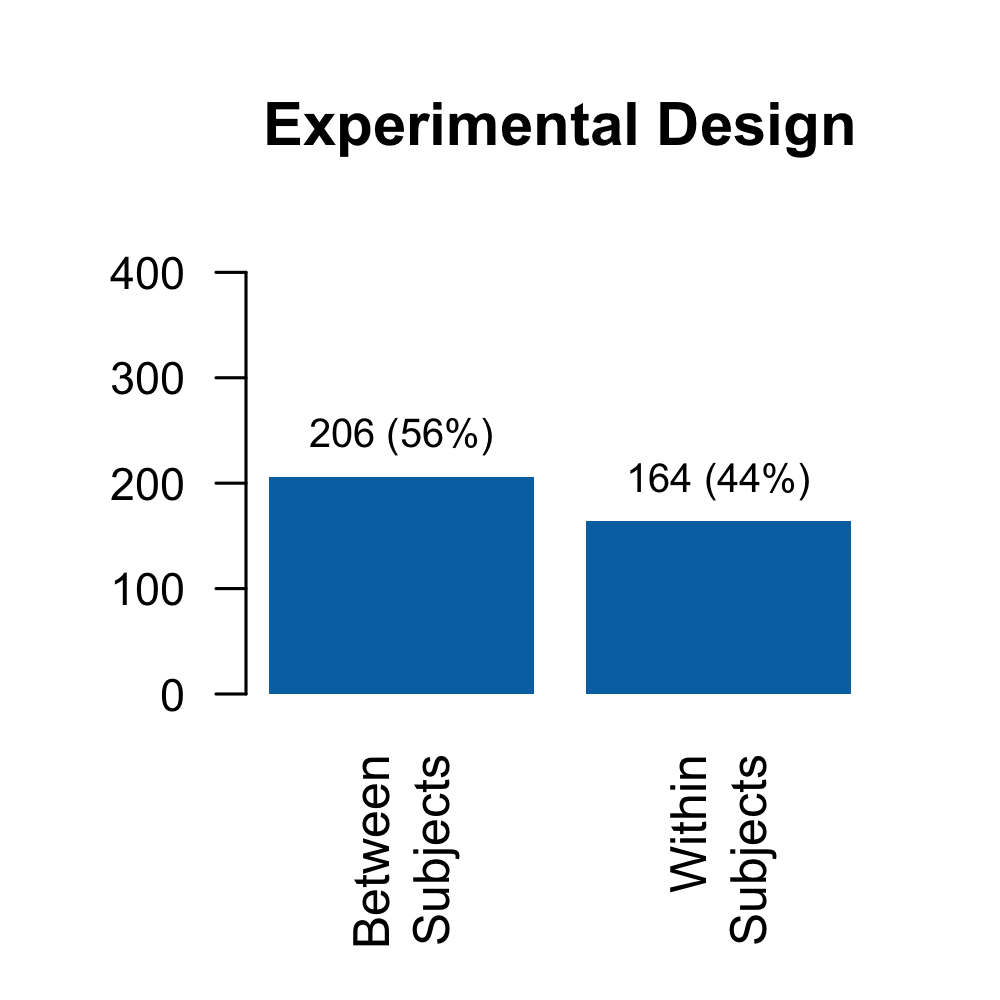}
\end{subfigure}
\hfill
\begin{subfigure}[t]{.32\textwidth}
    \centering
    \includegraphics[width=\linewidth]{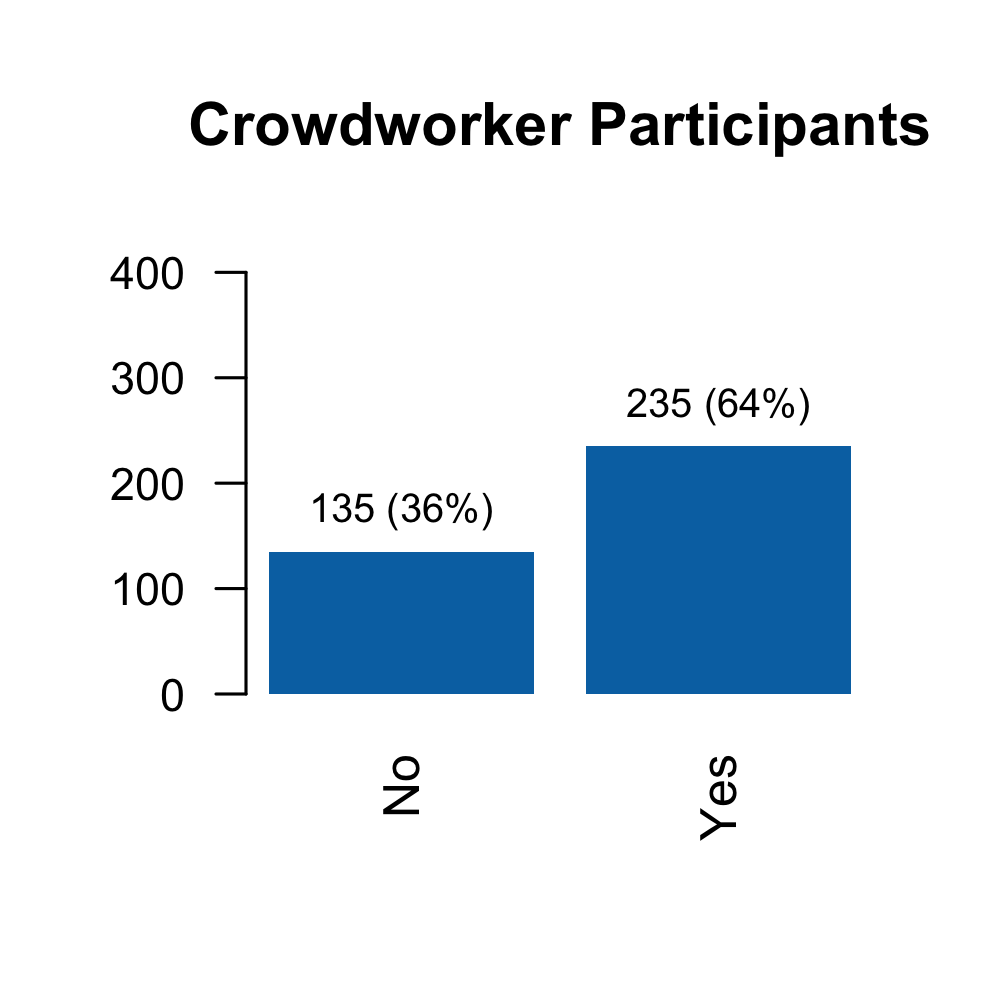}
\end{subfigure}
\hfill
\begin{subfigure}[t]{.32\textwidth}
    \centering
    \includegraphics[width=\linewidth]{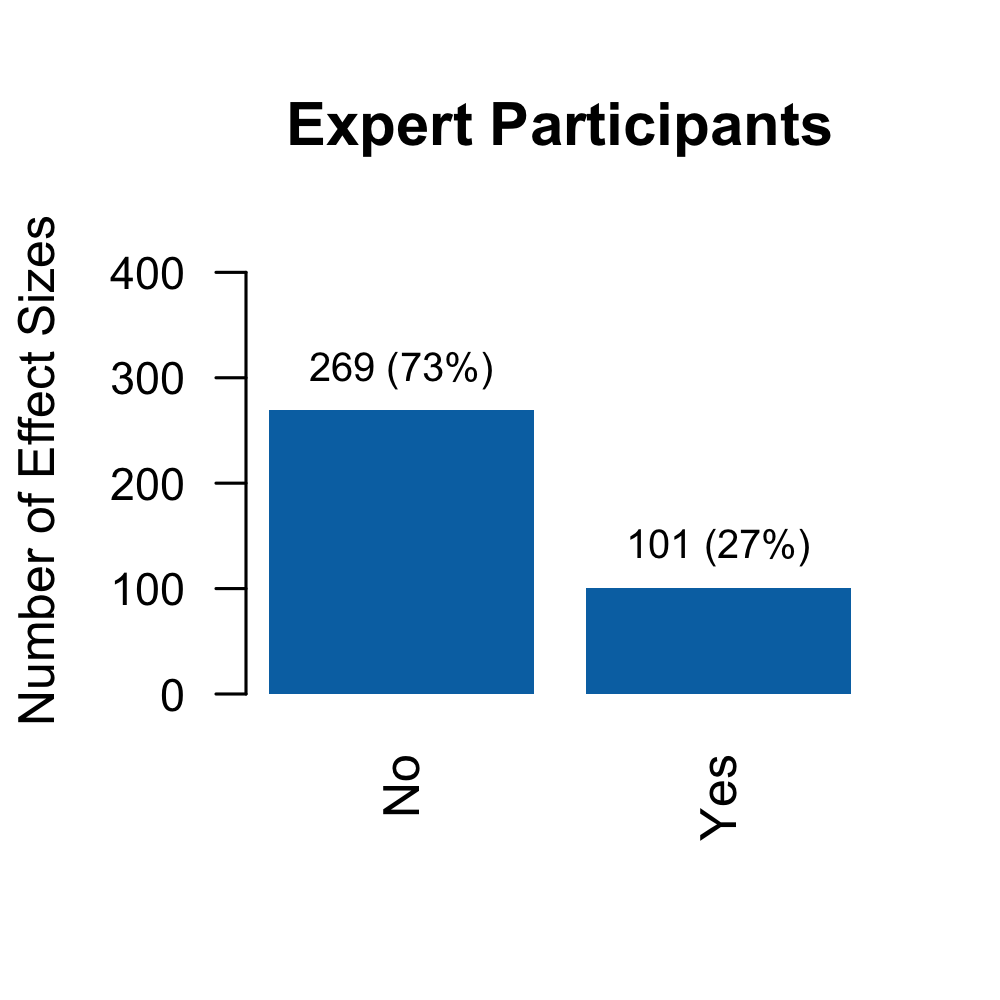}
\end{subfigure}
\begin{subfigure}[t]{.32\textwidth}
    \centering
    \includegraphics[width=\linewidth]{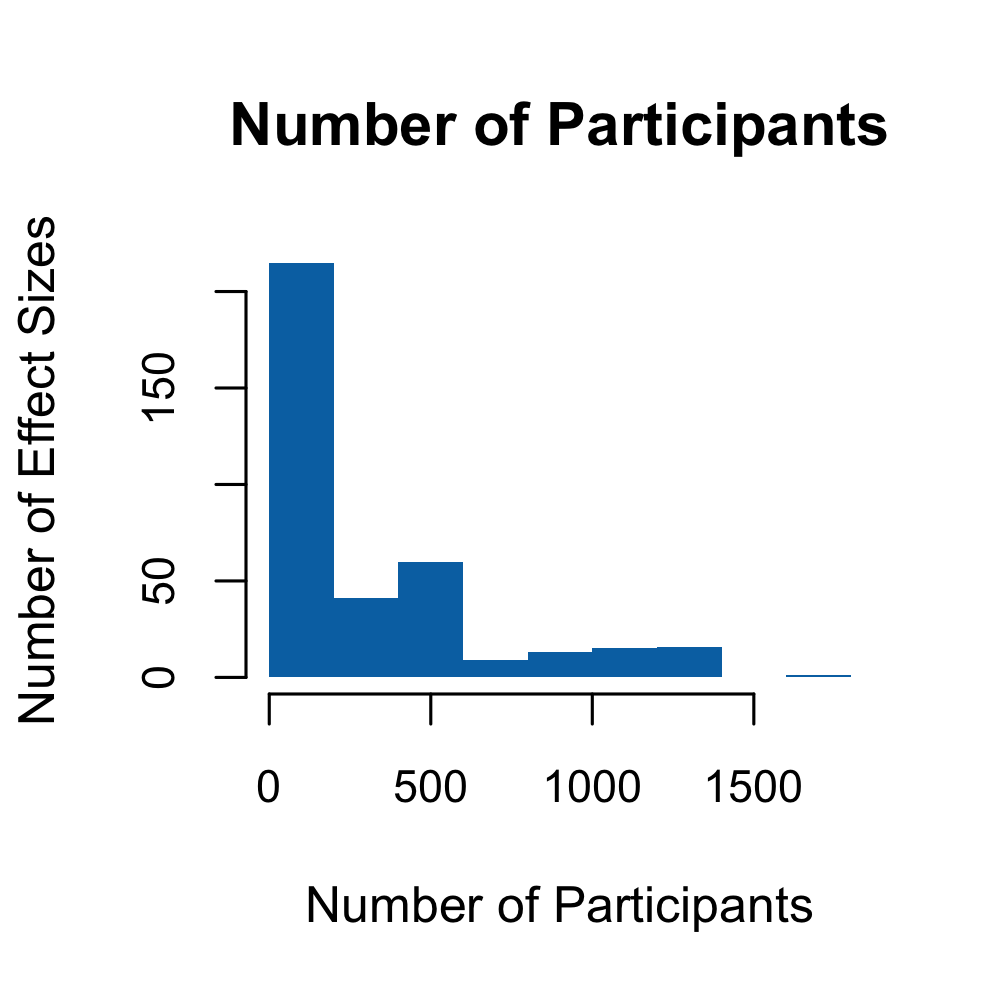}
\end{subfigure}
\hfill
\caption{Descriptive statistics for the effect sizes in our analysis.}
\label{fig:barplots}
\end{figure}

Figure \ref{fig:barplots} provides the key descriptive statistics of the effect sizes in our data set.  To highlight a few key takeaways, the vast majority come from experiments that evaluated human-AI performance in the context of a decision task, which often involves participants choosing between a set of options or providing a numerical value.  Far fewer come from creation tasks, which entail some kind of open-ended response ($n=34$, 9\%).  An even smaller number of effect sizes ($n=4$, 1\%) involve a pre-defined division of labor between the humans and AI tools, despite the potential for this process to lead to performance improvements.

\subsection{Effect Sizes for Additional Outcomes}
\begin{table}[H]
\centering
\footnotesize
\begin{tabular}{ccccc}
\toprule
\centering \textbf{Measure} & \textbf{Human-AI Synergy} & \textbf{Human Augmentation} & \textbf{AI Augmentation} & \textbf{Negative Synergy}\\ 
\midrule
$n$ & 370 & 370 & 370 & 370 \\ 
Hedges' $g$ & -0.23 & 0.64 & 0.30 & 1.14 \\
95\% CI & {[}-0.39, -0.07] & {[}0.53, 0.74] & {[}-0.03, 0.62] & {[}0.90, 1.38] \\
$t$-statistic & $t(92.39)=-2.89$ & $t(98.31)=11.88$  & $t(99.19)=1.82$ & $t(101.12)=9.58$ \\
$p$ & 0.005 & 0.000 & 0.072 & 0.000 \\ 
\bottomrule
\end{tabular}
    \caption{We fit separate meta-analytic models on the full set of results, and we report the results for human-AI synergy, human augmentation, AI augmentation, and negative synergy in this table. Here, $n$ corresponds to the number of effect sizes in the model, Hedges' $g$ is the meta-analytic average effect size, 95\% CI refers to the 95\% confidence interval for the effect size, $t$-statistic is the test statistic with the degrees of freedom estimated according to the Satterthwaite approximation, and $p$ is the two-sided $p$-value that indicates if the effect size is statistically significant from zero.}
    \label{tab:results_table}
\end{table}

\begin{table}[H]
\centering
\footnotesize
\begin{tabular}{ccccc} 
\toprule
\multicolumn{1}{l}{} & \multicolumn{2}{c}{\textbf{Human-AI Synergy}} & \multicolumn{2}{c}{\textbf{Human Augmentation}}  \\ 
\hline
Better Alone           & AI               & Human              & AI             & Human                   \\ 
\hline
$n$                      & 249              & 121                & 249            & 121                     \\
Hedges' $g$              & -0.53            & 0.39               & 0.73           & 0.39                    \\
95\% CI                & {[}-0.71, -0.35] & {[}0.21, 0.58]     & {[}0.62, 0.84] & {[}0.20, 0.58]          \\
$t$-statistic                & $t(69.73)=-5.79$        & $t(36.33)=4.27$          & $t(79.38)=12.81$      & $t(36.39)=4.23$              \\
$p$                & $0.000$        & $0.000$          & $0.000$      & $0.000$               \\
\bottomrule
\end{tabular}
    \caption{We fit separate meta-analytic models on the subset of results where (1) the AI performs better alone and (2) the human performs better alone, and we report the results for human-AI synergy and human augmentation in this table. Here, $n$ corresponds to the number of effect sizes in the model, Hedges' $g$ is the meta-analytic average effect size, 95\% CI refers to the 95\% confidence interval for the effect size, $t$-statistic is the test statistic with the degrees of freedom estimated according to the Satterthwaite approximation, and $p$ is the two-sided $p$-value that indicates if the effect size is statistically significant from zero.}
    \label{tab:results_table_2}
\end{table}

\begin{figure}[H]
  \begin{subfigure}[b]{0.45\textwidth}
    \includegraphics[width=\textwidth]{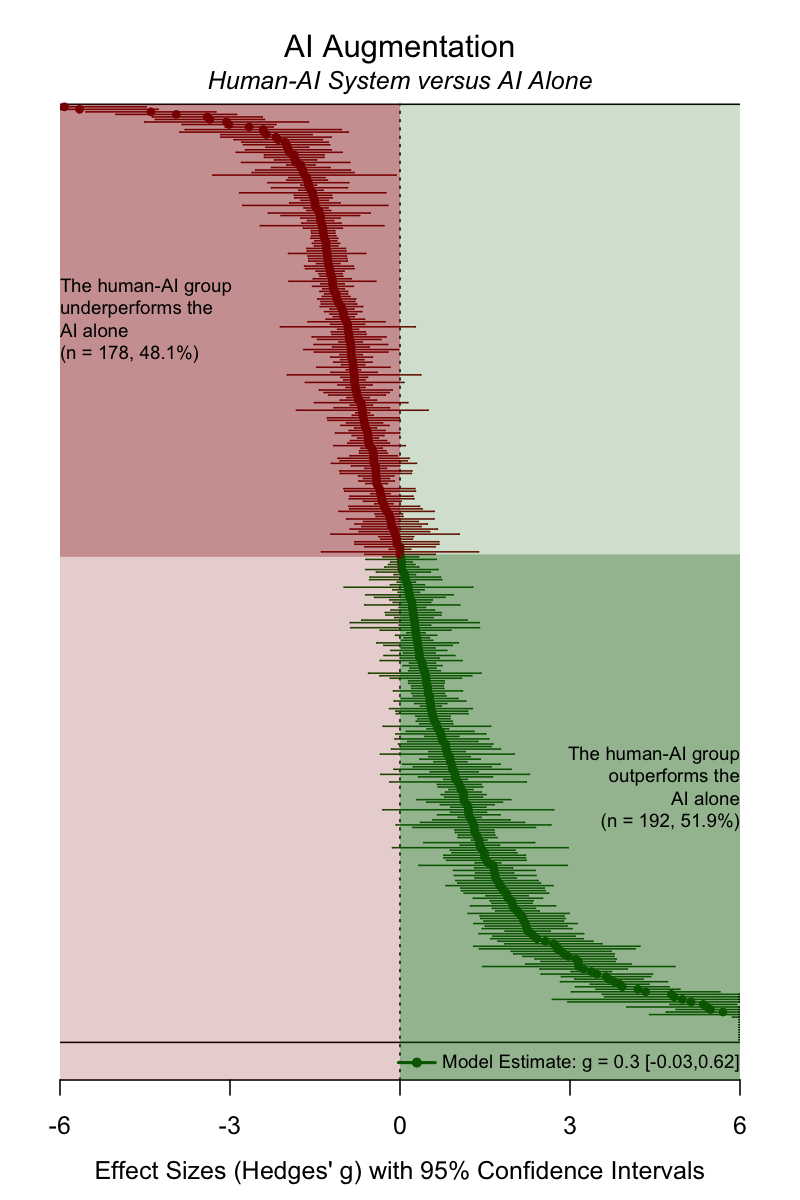}
    \label{fig:forest_plot_ai}
  \end{subfigure}
    \hfill
  \begin{subfigure}[b]{0.45\textwidth}
    \includegraphics[width=\textwidth]{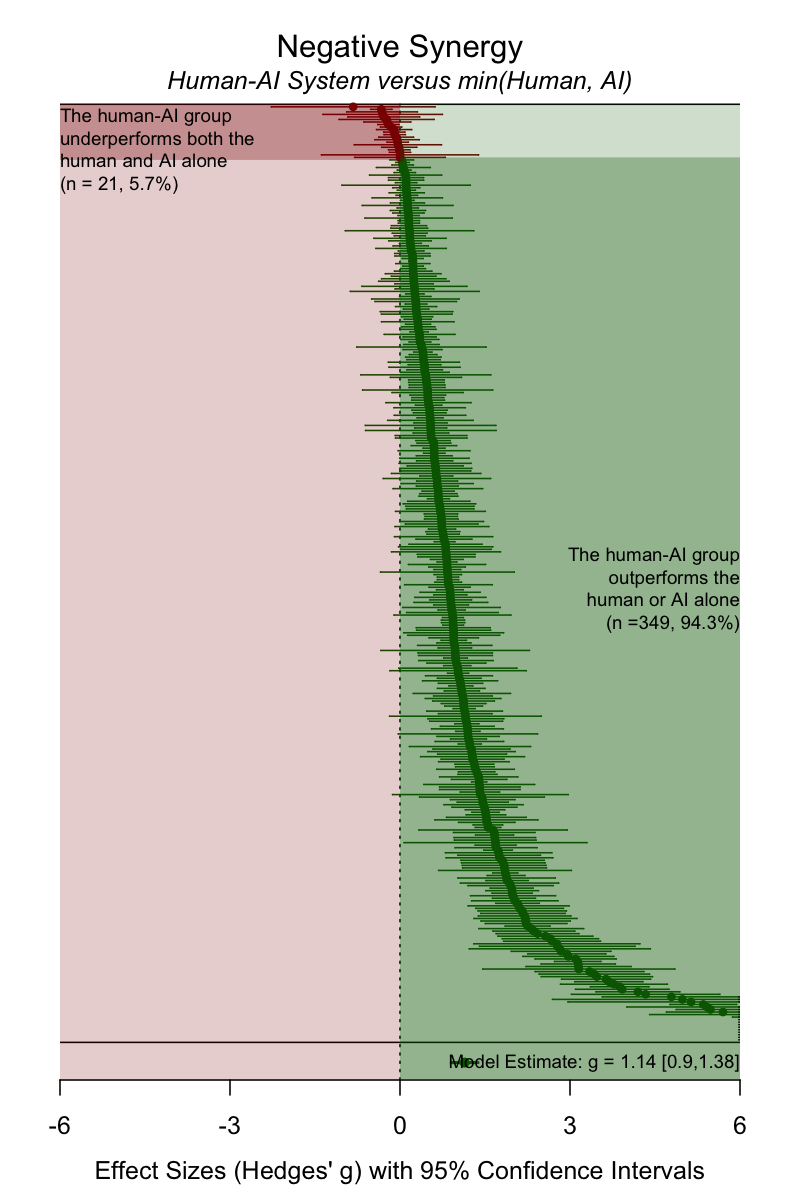}
    \label{fig:forest_plot_neg}
  \end{subfigure}
  
    \caption{Forest plots of all effect sizes ($n=370$) included in the meta-analysis for AI augmentation and negative synergy.  The positions of the points on the $x$-axes represent the values of the effect sizes, and the bars indicate the 95\% confidence interval for the effect sizes.  The colors of the points and lines correspond to the values of the effect sizes, with negative effect sizes (no human-AI synergy) colored red and positive effect sizes colored green (human-AI synergy).  The black dotted line corresponds to an effect size of Hedges' $g = 0$, which means that the human-AI group performed the same as the baseline.  The circle at the bottom of the graph represents the meta-analytic average effect size and confidence interval.}
    \label{fig:ForestPlot_Synergy_Other}
\end{figure}

\subsection{Heterogeneity Analyses}

\begin{table}[H]
\centering
\footnotesize
\begin{tabular}{ccccc}
\toprule
\textbf{Measure} & \textbf{Human-AI Synergy} & \textbf{Human Augmentation} & \textbf{AI Augmentation} & \textbf{Negative Synergy} \\ 
\midrule
$n$ & 370 & 370 & 370 & 370 \\ 
$\tau^2$ & 1.19 & 0.40 & 4.25 & 2.02 \\ 
$I^2$ & 97.7\% & 93.8\% & 99.3\% & 98.6\% \\ 
$I^2_{(2)}$ & 23.9\% & 43.6\% & 40.7\% & 50.1\% \\ 
$I^2_{(3)}$ & 73.8\% & 50.2\% & 58.6\% & 48.5\% \\ 
$Q$-Statistic & $Q(369)=8532$ & $Q(369)=4226$ & $Q(369) = 16017$ & $Q(369) = 7209$\\ 
$p$ & $0.000$ & $0.000$ & $0.000$ & $0.000$ \\ 
\bottomrule
\end{tabular}
    \caption{We fit separate meta-analytic models on the full set of results, and we report the heterogeneity metrics for the models of human-AI synergy, human augmentation, AI augmentation, and negative synergy. In particular, we provide $n$ (number of effect sizes), $\tau^2$, $I^2$, $I^2_{(2)}$, $I^2_{(3)}$, $Q$-statistic (with the corresponding degrees of freedom), and $p$ is the two-sided $p$-value from the $Q$-Test.}
    \label{tab:het_table}
\end{table}

\begin{table}[H]
\centering
\footnotesize
\begin{tabular}{ccccc} 
\toprule
\multicolumn{1}{l}{} & \multicolumn{2}{c}{\textbf{Human-AI Synergy}} & \multicolumn{2}{c}{\textbf{Human Augmentation}}  \\ 
\hline
Better Alone           & AI               & Human              & AI             & Human                   \\ 
\hline
$n$                      & 249              & 121                & 249            & 121                     \\
$\tau^2$            & 1.25             & 0.49               & 0.31           & 0.49                    \\
$I^2$              & 98.1\%          & 92.7\%            & 93.0\%        & 93.2\%                 \\
$I^2_{(2)}$           & 17.9\%          & 30.2\%            & 52.5\%        & 30.7\%                 \\
$I^2_{(3)}$           & 80.2\%          & 62.5\%            & 40.5\%        & 62.6\%                 \\
$Q$-statistic                 & $Q(248)=6111$        & $Q(120)=1071$          & $Q(248) = 2741$      & $Q(120) = 1136$               \\
$p$                 & $0.000$        & $0.000$          & $0.000$      & $0.000$               \\
\bottomrule
\end{tabular}
    \caption{We fit separate meta-analytic models on the subset of results where (1) the AI performs better alone and (2) the human performs better alone, and we report the heterogeneity metrics for the models of human-AI synergy and human augmentation. In particular, we provide $n$ (number of effect sizes), $\tau^2$, $I^2$, $I^2_{(2)}$, $I^2_{(3)}$, $Q$-statistic (with the corresponding degrees of freedom), and $p$ is the two-sided $p$-value from the $Q$-Test.}
    \label{tab:het_table_2}
\end{table}

\subsection{Moderators for Additional Outcomes}

\begin{table}[H]
    \centering
    \includegraphics{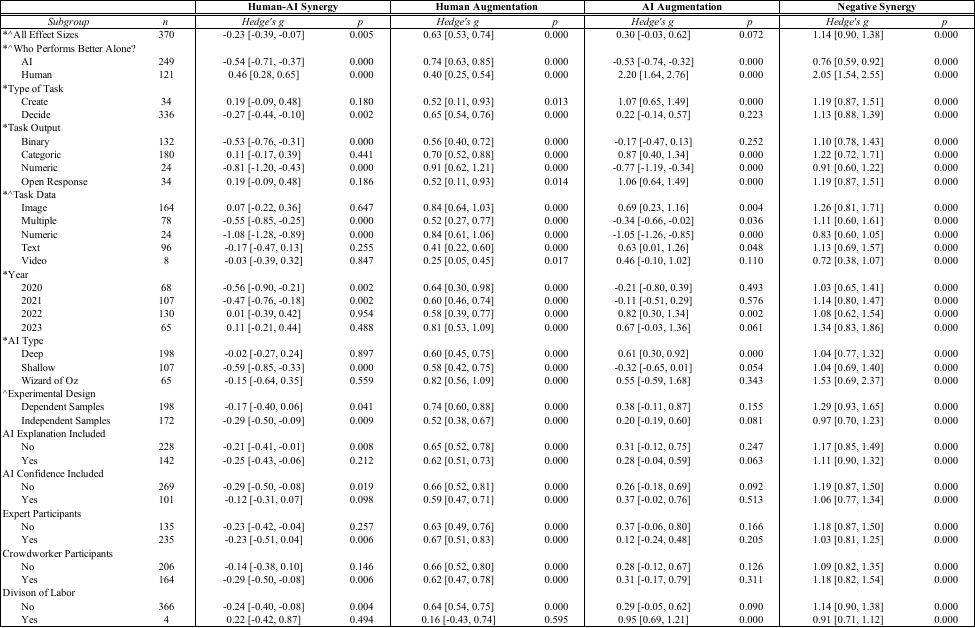}
    \caption{Results from the three-level meta-regression models for the moderator variables.  Here, $n$ is the number of included effect sizes for the moderator subgroup level, $g$ is the estimated effect size with the corresponding 95\% confidence interval for the subgroup level, and $p$ is the two-sided $p$-value of the estimated effect size for the subgroup level, which tests if the effect size is  significantly different from zero.}
    \label{tab:mod_table}
\end{table}


\subsection{Scatterplots of Effect Sizes for Accuracy-Based Tasks}

\begin{figure}[H]
\centering
\includegraphics[width=\linewidth]{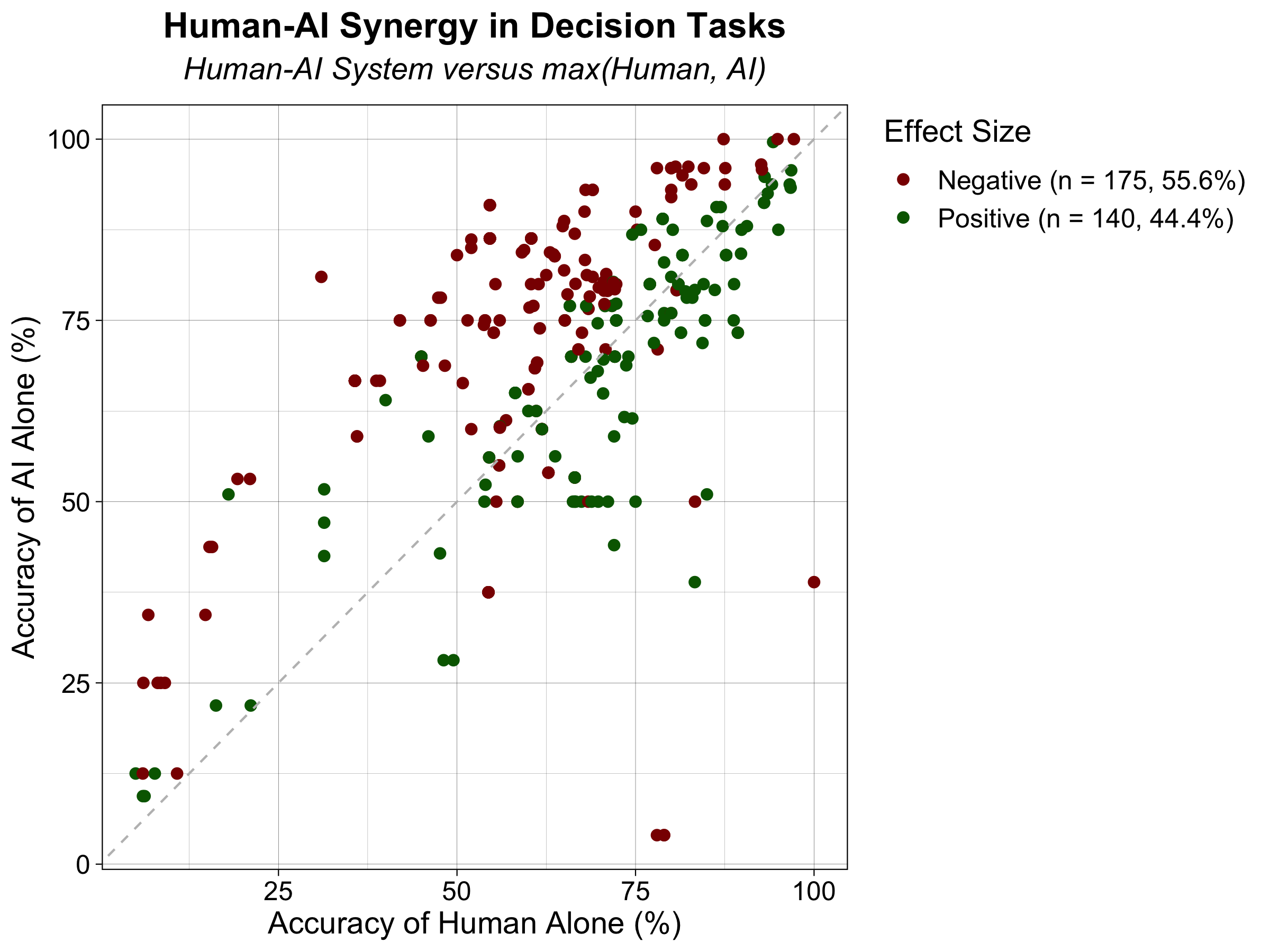}
\caption{Scatterplot of the effect sizes that correspond to decision tasks where the researchers measured performance according to accuracy.  The position of the points on the $x$-axis represents the accuracy of the human alone and the position of the points on the $y$-axis represents the accuracy of the AI alone.  The colors of the points correspond to the value of the effect sizes, with negative effect sizes red and positive effect sizes green.  The gray dotted line indicates that the accuracy of the humans alone and AI alone are equivalent.  In the region above this line, the accuracy of the AI alone exceeds that of the human alone, and we observe a large proportion of red points -- no evidence of human-AI synergy.  In the region below this line, the accuracy of the humans alone exceeds that of the AI alone, and we observe a large proportion of green points -- evidence human-AI synergy.}
\label{fig:acc_synergy}
\end{figure}

\begin{figure}[H]
\centering
\includegraphics[width=\linewidth]{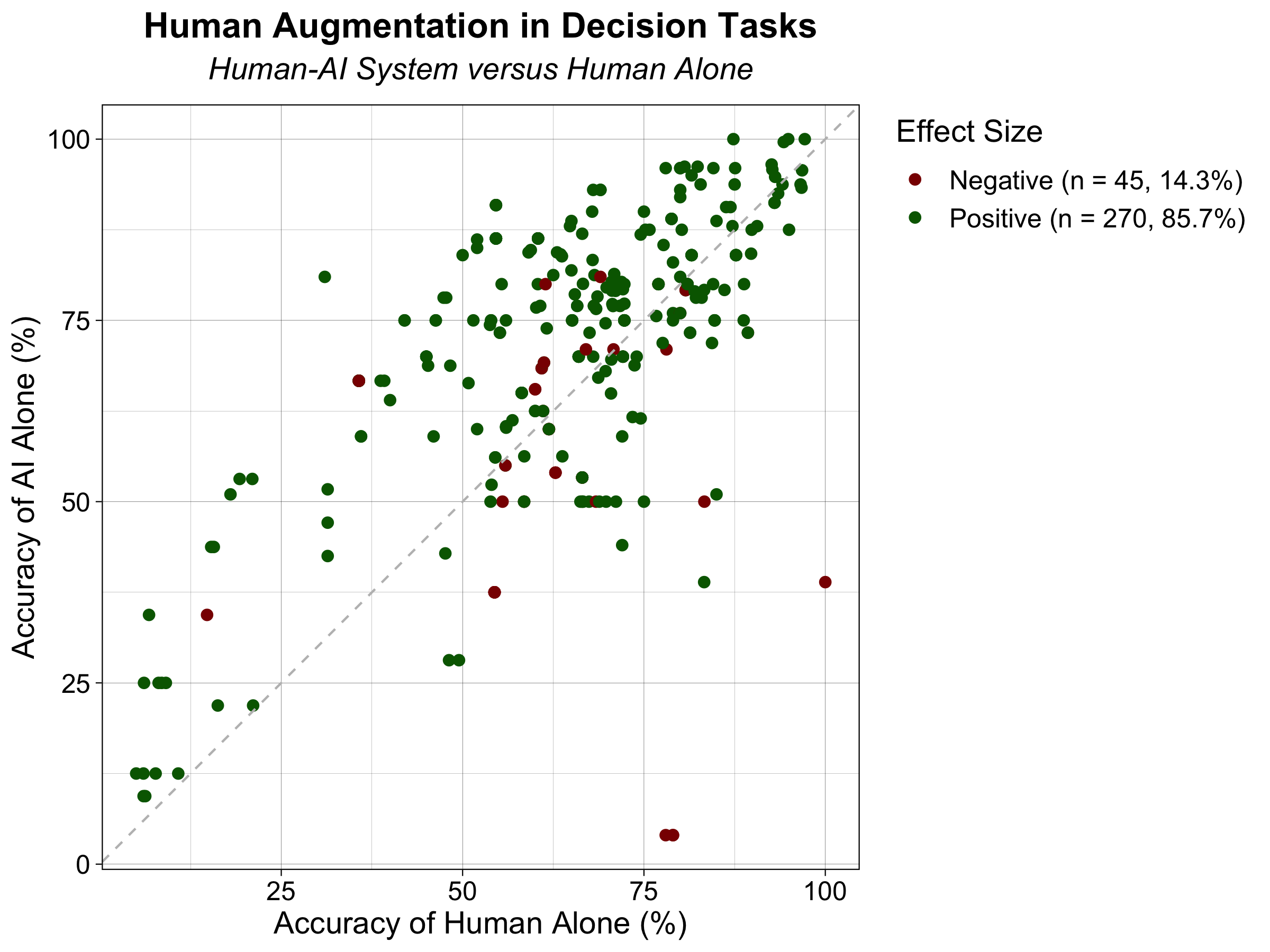}
\caption{Scatterplot of the effect sizes that correspond to decision tasks where the researchers measured performance according to accuracy.  The position of the points on the $x$-axis represents the accuracy of the human alone and the position of the points on the $y$-axis represents the accuracy of the AI alone.  The colors of the points correspond to the value of the effect sizes, with negative effect sizes red and positive effect sizes green.  The gray dotted line indicates that the accuracy of the humans alone and AI alone are equivalent.}
\label{fig:acc_synergy_h}
\end{figure}

\begin{figure}[H]
\centering
\includegraphics[width=\linewidth]{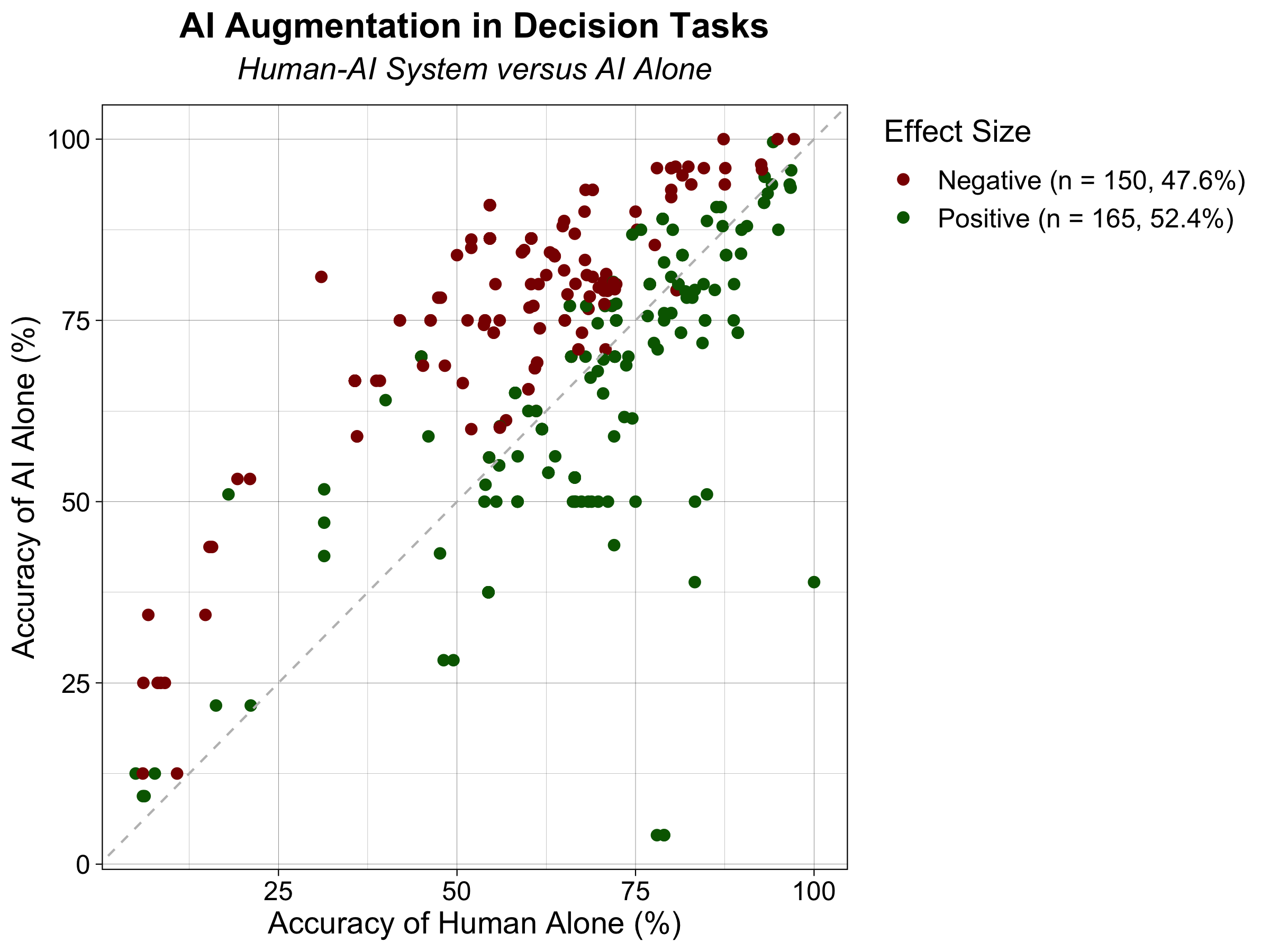}
\caption{Scatterplot of the effect sizes that correspond to decision tasks where the researchers measured performance according to accuracy.  The position of the points on the $x$-axis represents the accuracy of the human alone and the position of the points on the $y$-axis represents the accuracy of the AI alone.  The colors of the points correspond to the value of the effect sizes, with negative effect sizes red and positive effect sizes green.  The gray dotted line indicates that the accuracy of the humans alone and AI alone are equivalent.}
\label{fig:acc_synergy_a}
\end{figure}

\begin{figure}[H]
\centering
\includegraphics[width=\linewidth]{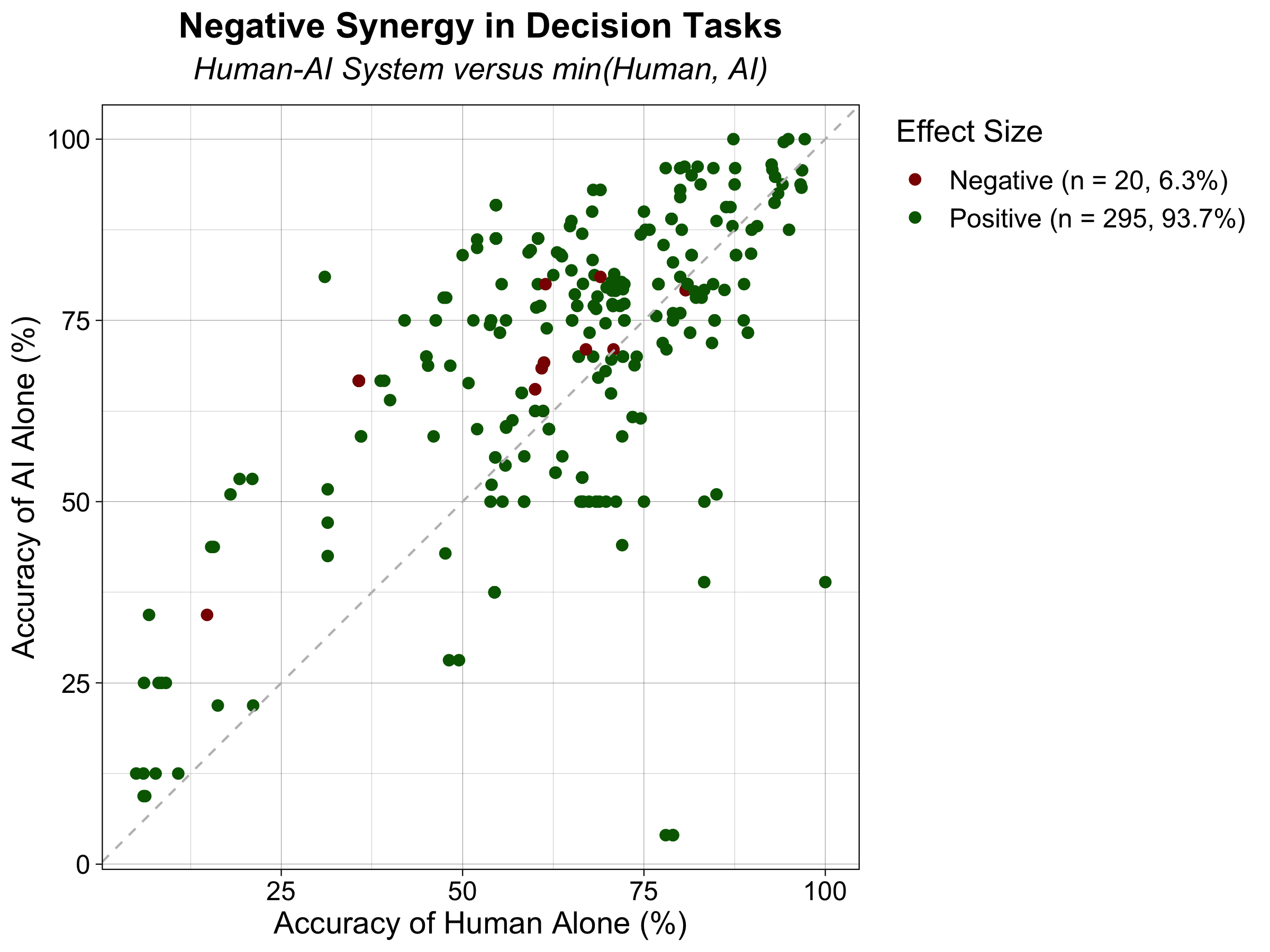}
\caption{Scatterplot of the effect sizes that correspond to decision tasks where the researchers measured performance according to accuracy.  The position of the points on the $x$-axis represents the accuracy of the human alone and the position of the points on the $y$-axis represents the accuracy of the AI alone.  The colors of the points correspond to the value of the effect sizes, with negative effect sizes red and positive effect sizes green.  The gray dotted line indicates that the accuracy of the humans alone and AI alone are equivalent.}
\label{fig:acc_synergy_n}
\end{figure}

\subsection{Division of Labor}\label{div_labor}
Only 3 of the 100+ experiments in our analysis explore processes with a pre-determined delegation of separate sub-tasks to humans and AI, so the difference between effect sizes with and without division of labor was not statistically significant. However, with the 4 effect sizes from these 3 experiments, we did find that positive human-AI synergy ($g = 0.22$, $t(104)=0.69$, two-tailed $p=0.494$, 95\% CI $-0.42$ to $0.87$) occurred among experiments with a predetermined division of labor between the human and AI, while in the experiments without this feature, the effect size was significantly negative ($g = -0.24$, $t(104)=-2.93$, two-tailed $p=0.004$, 95\% CI $-0.40$ to $-0.08$).  For example, in Lee et al. (2021), physical therapists review patient-specific analysis and then provide feature-based feedback to an AI-system, which then generates an assessment for that patient based on its training data as well as this new information \cite{lee2021a}.  Here, the authors find that their human-AI system achieves greater accuracy (93\%) than either the therapist alone (75\%) or AI alone without the therapist input (87\%).  

Hemmer at al. (2023) employ a different division of labor in an experiment involving image classification \cite{hemmer2023human}. They develop an AI model that learns to both classify images and estimate human classifications of the same image.  Then, for each image, this model classifies the images for which it has a higher confidence than it predicts the human would have and assigns the rest of the images to a human.  These authors find that the human-AI system outperforms the human alone and AI alone (84\% accuracy versus 67\% and 75\%, respectively).  

Lastly, Lai et al. (2022) also create a pre-determined division of labor where an AI system generates a summary of a post, which a human can edit and refine \cite{lai2022exploration}.  They find, however, that the human alone produces higher quality summaries (5.8/7) than both the AI alone (4.7/7) and the human-AI combination (5.5/7).  So this particular task allocation does not lead to human-AI synergy.

\subsection{Effect Sizes over Time}\label{variation_by_time}
\begin{figure}[H]
\centering
\begin{subfigure}[t]{.8\textwidth}
    \centering
    \includegraphics[width=\linewidth]{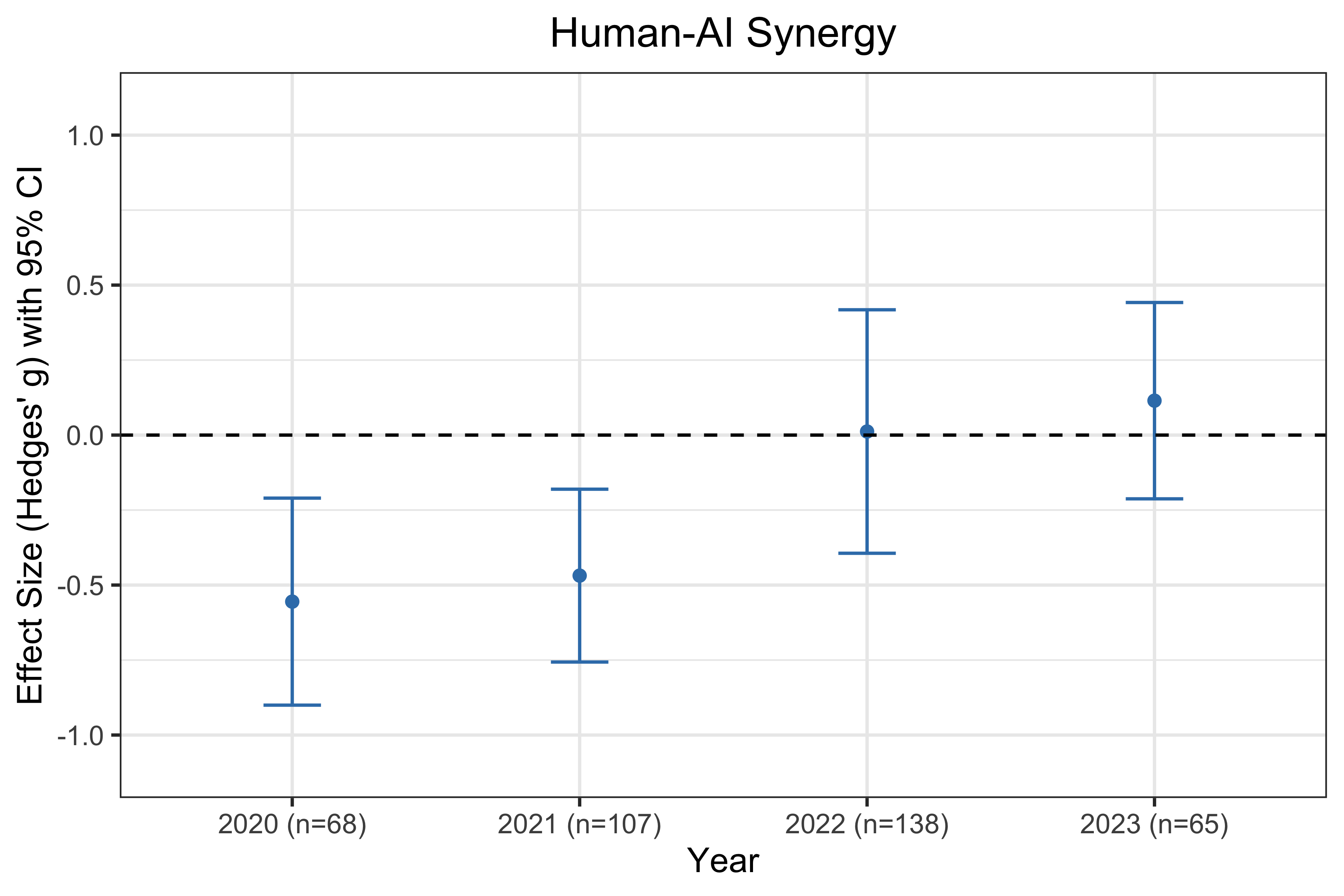}
    \label{fig:Year_Synergy_S}
\end{subfigure}
\hfill
\begin{subfigure}[t]{.8\textwidth}
    \centering
    \includegraphics[width=\linewidth]{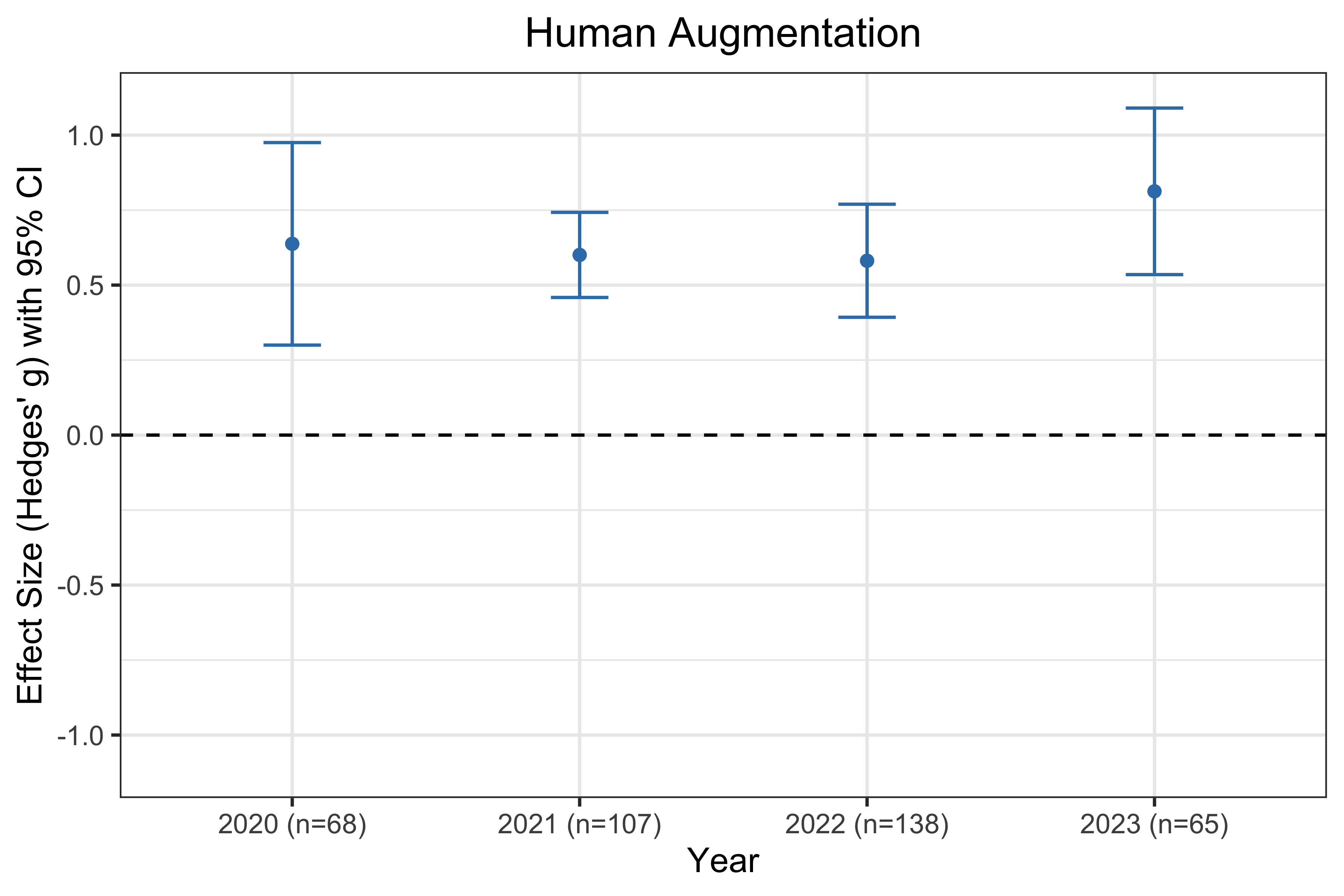}
    \label{fig:Year_Synergy_H}
\end{subfigure}
\hfill
\caption{Plots of the meta-analytic average effect sizes for human-AI synergy and human augmentation by year of publication. The error bars correspond to 95\% confidence intervals, and the dotted black line corresponds to an effect size of zero (no effect).}
\label{fig:year_plots}
\end{figure}

When we look at the evolution of human-AI synergy over the past years (see Figure \ref{fig:year_plots}), we observe potential signs of progress. 
This suggestive trend may reflect increased attention to designing studies that elicit human-AI synergy. Notably, however, the progress---if it exists---is not by any means rapid, and stands in stark contrast to the quick pace of development of AI systems themselves, particularly large language models (LLMs). We hope that our work can spur further progress in this domain.   

\subsection{Bias Tests}
\begin{figure}[H]
\centering
\begin{subfigure}[t]{.45\textwidth}
    \centering
    \includegraphics[width=\linewidth]{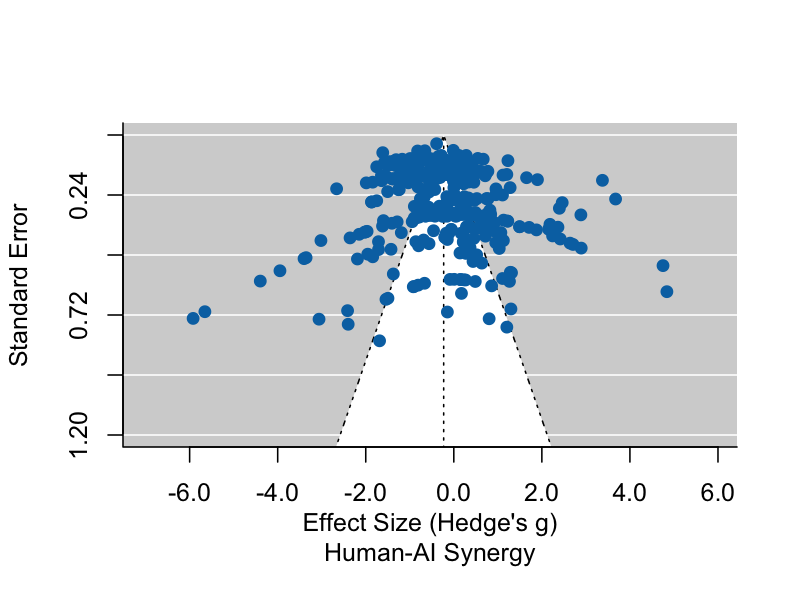}
    \label{fig:funnel_plot_strong}
\end{subfigure}
\hfill
\begin{subfigure}[t]{.45\textwidth}
    \centering
    \includegraphics[width=\linewidth]{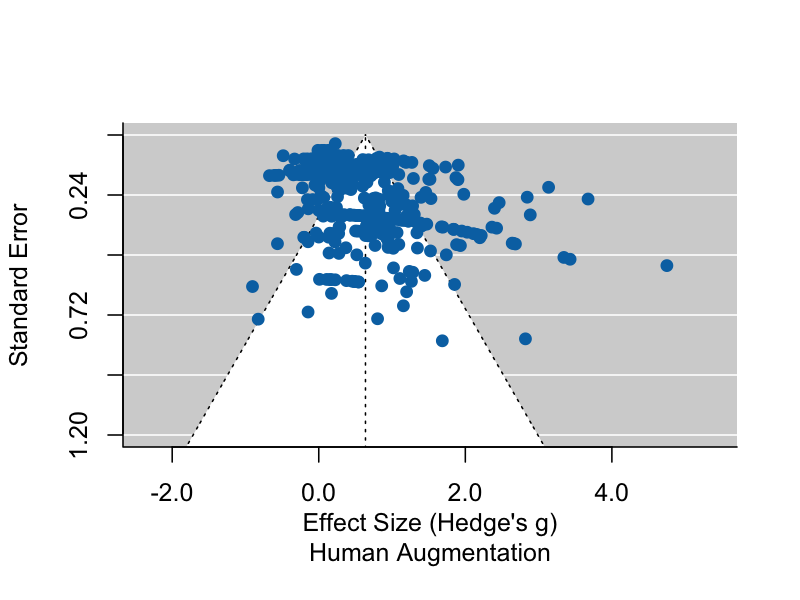}
    \label{fig:funnel_plot_human}
\end{subfigure}
\hfill
\begin{subfigure}[t]{.45\textwidth}
    \centering
    \includegraphics[width=\linewidth]{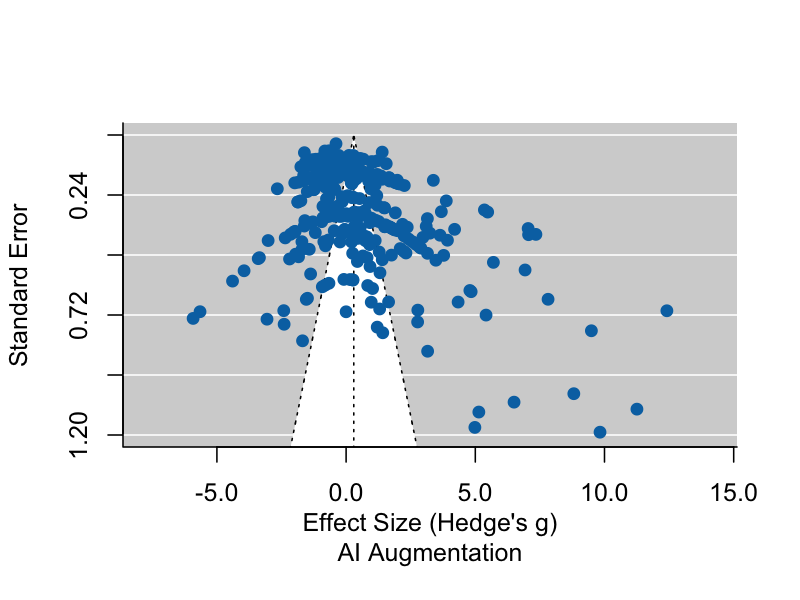}
    \label{fig:funnel_plot_ai}
\end{subfigure}
\hfill
\begin{subfigure}[t]{.45\textwidth}
    \centering
    \includegraphics[width=\linewidth]{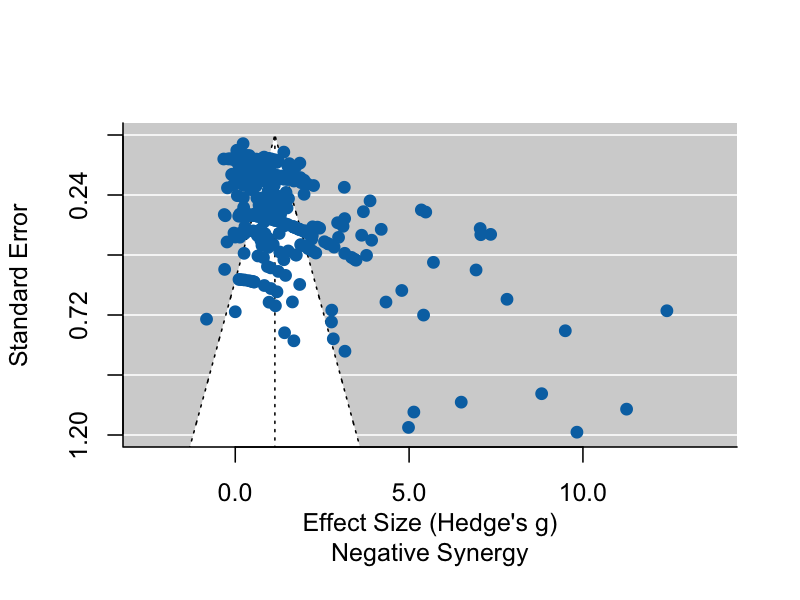}
    \label{fig:funnel_plot_neg}
\end{subfigure}
\hfill

\caption{Funnel plots of the observed effect sizes and corresponding standard errors for human-AI synergy, human augmentation, AI augmentation, and negative synergy.  The blue dots represent effect sizes and the vertical dotted black line represents the estimated pooled effect sizes from the meta-analytic models.  The area shaded in white represents regions of statistical non-significance ($p > 0.05$), and the area shaded in gray represents regions of statistical significance ($p \leq 0.05$).  In the absence of publication bias, we expect the points to fall roughly symmetrically around the estimated pooled effect size with a large proportion within the white triangle.  We observe this pattern in the plot for human-AI synergy, but less so in the plots for human augmentation, AI augmentation, and negative synergy.}
\label{fig:funnel_plots}
\end{figure}



\printbibliography[title={References}]
\end{refsection}

\end{document}